\documentclass[11pt]{article}
\usepackage[dvips]{graphicx}
\usepackage{epsfig}
\usepackage{amssymb}
\usepackage[left]{lineno}

\def\NIMA#1#2#3{Nucl. Inst. Methods {\bf A#1} (#2) #3}

\begin{document}
\centerline{\LARGE EUROPEAN ORGANIZATION FOR NUCLEAR RESEARCH}
\vspace{10mm} {\flushright{
CERN-PH-EP-2010-051 \\
22 November 2010
\\}}
\vspace{15mm}

\begin{center}
{\bf {\Large \boldmath{New measurement of the $K^\pm\to\pi^\pm
\mu^+\mu^-$ decay }}}
\end{center}
%\vspace{4mm}
\begin{center}
{\Large The NA48/2 Collaboration$\,$\renewcommand{\thefootnote}{\fnsymbol{footnote}}%
\footnotemark[1]\renewcommand{\thefootnote}{\arabic{footnote}}}\\
\end{center}

\begin{abstract}
A sample of 3120 $K^\pm\to\pi^\pm\mu^+\mu^-$ decay candidates with
$(3.3\pm0.7)\%$ background contamination has been collected by the
NA48/2 experiment at the CERN SPS, allowing a detailed study of the
decay properties. The branching ratio was measured to be ${\rm
BR}=(9.62\pm0.25)\times 10^{-8}$. The form factor $W(z)$, where
$z=(M_{\mu\mu}/M_K)^2$, was parameterized according to several
models. In particular, the slope of the linear form factor
$W(z)=W_0(1+\delta z)$ was measured to be $\delta=3.11\pm0.57$.
Upper limits of $2.9\times 10^{-2}$ and $2.3\times 10^{-2}$ on
possible charge asymmetry and forward-backward asymmetry were
established at $90\%$ CL. An upper limit ${\rm
BR}(K^\pm\to\pi^\mp\mu^\pm\mu^\pm)<1.1\times 10^{-9}$ was
established at $90\%$ CL for the rate of the lepton number violating
decay.
\end{abstract}

\begin{center}
\it{To be submitted for publication in Physics Letters B}
\end{center}

\setcounter{footnote}{0}
\renewcommand{\thefootnote}{\fnsymbol{footnote}}
\footnotetext[1]{Copyright CERN for the benefit of the
Collaboration.}
\renewcommand{\thefootnote}{\arabic{footnote}}

\newpage

\begin{center}
{\Large The NA48/2 Collaboration}\\
\vspace{2mm}
 J.R.~Batley,
 G.~Kalmus,
 C.~Lazzeroni$\,$\footnotemark[1],
 D.J.~Munday,
 M.W.~Slater$\,$\footnotemark[1],
 S.A.~Wotton \\
{\em \small Cavendish Laboratory, University of Cambridge,
Cambridge, CB3 0HE, UK$\,$\footnotemark[2]} \\[0.2cm]
 R.~Arcidiacono$\,$\footnotemark[3],
 G.~Bocquet,
 N.~Cabibbo$\,$\footnotemark[4],
 A.~Ceccucci,
 D.~Cundy$\,$\footnotemark[5],
 V.~Falaleev,
 M.~Fidecaro,
 L.~Gatignon,
 A.~Gonidec,
 W.~Kubischta,
 A.~Norton$\,$\footnotemark[6],
 A.~Maier,\\
 M.~Patel$\,$\footnotemark[7],
 A.~Peters\\
{\em \small CERN, CH-1211 Gen\`eve 23, Switzerland} \\[0.2cm]
 S.~Balev$\,$\footnotemark[8],
 P.L.~Frabetti,
 E.~Goudzovski$\,$\renewcommand{\thefootnote}{\fnsymbol{footnote}}%
\footnotemark[2]\renewcommand{\thefootnote}{\arabic{footnote}}\footnotemark[1],
 P.~Hristov$\,$\footnotemark[8],
 V.~Kekelidze,
 V.~Kozhuharov$\,$\footnotemark[9],
 L.~Litov,
 D.~Madigozhin,
 E.~Marinova$\,$\footnotemark[10],
 N.~Molokanova,
 I.~Polenkevich,\\
 Yu.~Potrebenikov,
 S.~Stoynev$\,$\footnotemark[11],
 A.~Zinchenko \\
{\em \small Joint Institute for Nuclear Research, 141980 Dubna,
Moscow region, Russia} \\[0.2cm]
 E.~Monnier$\,$\footnotemark[12],
 E.~Swallow,
 R.~Winston\\
{\em \small The Enrico Fermi Institute, The University of Chicago,
Chicago, IL 60126, USA}\\[0.2cm]
 P.~Rubin$\,$\footnotemark[13],
 A.~Walker \\
{\em \small Department of Physics and Astronomy, University of
Edinburgh, JCMB King's Buildings, Mayfield Road, Edinburgh, EH9 3JZ, UK} \\[0.2cm]
 W.~Baldini,
 A.~Cotta Ramusino,
 P.~Dalpiaz,
 C.~Damiani,
 M.~Fiorini$\,$\footnotemark[8],
 A.~Gianoli,
 M.~Martini,
 F.~Petrucci,
 M.~Savri\'e,
 M.~Scarpa,
 H.~Wahl \\
{\em \small Dipartimento di Fisica dell'Universit\`a e Sezione
dell'INFN di Ferrara, I-44100 Ferrara, Italy} \\[0.2cm]
 A.~Bizzeti$\,$\footnotemark[14],
 M.~Lenti,
 M.~Veltri$\,$\footnotemark[15] \\
{\em \small Sezione dell'INFN di Firenze, I-50125 Firenze, Italy} \\[0.2cm]
 M.~Calvetti,
 E.~Celeghini,
 E.~Iacopini,
 G.~Ruggiero$\,$\footnotemark[16] \\
{\em \small Dipartimento di Fisica dell'Universit\`a e Sezione
dell'INFN di Firenze, I-50125 Firenze, Italy} \\[0.2cm]
 M.~Behler,
 K.~Eppard,
 K.~Kleinknecht,
 P.~Marouelli,
 L.~Masetti,
 U.~Moosbrugger,
 C.~Morales Morales,
 B.~Renk,
 M.~Wache,
 R.~Wanke,
 A.~Winhart \\
{\em \small Institut f\"ur Physik, Universit\"at Mainz, D-55099
 Mainz, Germany$\,$\footnotemark[17]} \\[0.2cm]
 D.~Coward$\,$\footnotemark[18],
 A.~Dabrowski$\,$\footnotemark[8],
 T.~Fonseca Martin$\,$\footnotemark[19],
 M.~Shieh,
 M.~Szleper,\\
 M.~Velasco,
 M.D.~Wood$\,$\footnotemark[20] \\
{\em \small Department of Physics and Astronomy, Northwestern
University, Evanston, IL 60208, USA}\\[0.2cm]
 P.~Cenci,
 M.~Pepe,
 M.C.~Petrucci \\
{\em \small Sezione dell'INFN di Perugia, I-06100 Perugia, Italy} \\[0.2cm]
 G.~Anzivino,
 E.~Imbergamo,
 A.~Nappi,
 M.~Piccini,
 M.~Raggi$\,$\footnotemark[21],
 M.~Valdata-Nappi \\
{\em \small Dipartimento di Fisica dell'Universit\`a e
Sezione dell'INFN di Perugia, I-06100 Perugia, Italy} \\[0.2cm]
 C.~Cerri,
 R.~Fantechi \\
{\em Sezione dell'INFN di Pisa, I-56100 Pisa, Italy} \\[0.2cm]
 G.~Collazuol,
 L.~DiLella,
 G.~Lamanna$\,$\footnotemark[8],
 I.~Mannelli,
 A.~Michetti \\
{\em Scuola Normale Superiore e Sezione dell'INFN di Pisa, I-56100
Pisa, Italy} \\[0.2cm]
 F.~Costantini,
 N.~Doble,
 L.~Fiorini$\,$\footnotemark[22],
 S.~Giudici,
 G.~Pierazzini,\
 M.~Sozzi,
 S.~Venditti \\
{\em Dipartimento di Fisica dell'Universit\`a e Sezione dell'INFN di
Pisa, I-56100 Pisa, Italy} \\[0.2cm]
 B.~Bloch-Devaux$\,$\footnotemark[23],
 C.~Cheshkov$\,$\footnotemark[24],
 J.B.~Ch\`eze,
 M.~De Beer,
 J.~Derr\'e,
 G.~Marel,
 E.~Mazzucato,
 B.~Peyaud,
 B.~Vallage \\
{\em \small DSM/IRFU -- CEA Saclay, F-91191 Gif-sur-Yvette, France} \\[0.2cm]
\newpage
 M.~Holder,
 M.~Ziolkowski \\
{\em \small Fachbereich Physik, Universit\"at Siegen, D-57068
 Siegen, Germany$\,$\footnotemark[25]} \\[0.2cm]
 C.~Biino,
 N.~Cartiglia,
 F.~Marchetto \\
{\em \small Sezione dell'INFN di Torino, I-10125 Torino, Italy} \\[0.2cm]
 S.~Bifani$\,$\footnotemark[26],
 M.~Clemencic$\,$\footnotemark[8],
 S.~Goy Lopez$\,$\footnotemark[27] \\
{\em \small Dipartimento di Fisica Sperimentale dell'Universit\`a e
Sezione dell'INFN di Torino,\\ I-10125 Torino, Italy} \\[0.2cm]
 H.~Dibon,
 M.~Jeitler,
 M.~Markytan,
 I.~Mikulec,
 G.~Neuhofer,
 L.~Widhalm \\
{\em \small \"Osterreichische Akademie der Wissenschaften, Institut
f\"ur Hochenergiephysik,\\ A-10560 Wien, Austria$\,$\footnotemark[28]} \\[0.5cm]
\end{center}

\setcounter{footnote}{0}
\renewcommand{\thefootnote}{\fnsymbol{footnote}}
\footnotetext[2]{Corresponding author, email:
eg@hep.ph.bham.ac.uk}
\renewcommand{\thefootnote}{\arabic{footnote}}
\footnotetext[1]{University of Birmingham, Edgbaston, Birmingham,
B15 2TT, UK}
\footnotetext[2]{Funded by the UK Particle Physics and Astronomy
Research Council}
\footnotetext[3]{Dipartimento di Fisica Sperimentale
dell'Universit\`a e Sezione dell'INFN di Torino, I-10125 Torino,
Italy}
\footnotetext[4]{Universit\`a di Roma ``La Sapienza'' e Sezione
dell'INFN di Roma, I-00185 Roma, Italy}
\footnotetext[5]{Istituto di Cosmogeofisica del CNR di Torino,
I-10133 Torino, Italy}
\footnotetext[6]{Dipartimento di Fisica dell'Universit\`a e Sezione
dell'INFN di Ferrara, I-44100 Ferrara, Italy}
\footnotetext[7]{Department of Physics, Imperial College, London,
SW7 2BW, UK}
\footnotetext[8]{CERN, CH-1211 Gen\`eve 23, Switzerland}
\footnotetext[9]{Faculty of Physics, University of Sofia ``St. Kl.
Ohridski'', 5 J. Bourchier Blvd., 1164 Sofia, Bulgaria}
\footnotetext[10]{Sezione dell'INFN di Perugia, I-06100 Perugia,
Italy}
\footnotetext[11]{Northwestern University, 2145 Sheridan Road,
Evanston, IL 60208, USA}
\footnotetext[12]{Centre de Physique des Particules de Marseille,
IN2P3-CNRS, Universit\'e de la M\'editerran\'ee, F-13288 Marseille,
France}
\footnotetext[13]{Department of Physics and Astronomy, George Mason
University, Fairfax, VA 22030, USA}
\footnotetext[14]{Dipartimento di Fisica, Universit\`a di Modena e
Reggio Emilia, I-41100 Modena, Italy}
\footnotetext[15]{Istituto di Fisica, Universit\`a di Urbino,
I-61029 Urbino, Italy}
\footnotetext[16]{Scuola Normale Superiore, I-56100 Pisa, Italy}
\footnotetext[17]{Funded by the German Federal Minister for
Education and research under contract 05HK1UM1/1}
\footnotetext[18]{SLAC, Stanford University, Menlo Park, CA 94025,
USA}
\footnotetext[19]{Laboratory for High Energy Physics, CH-3012 Bern,
Switzerland}
\footnotetext[20]{UCLA, Los Angeles, CA 90024, USA}
\footnotetext[21]{Laboratori Nazionali di Frascati, via E. Fermi,
40, I-00044 Frascati (Rome), Italy}
\footnotetext[22]{Institut de F\'isica d'Altes Energies, UAB,
E-08193 Bellaterra (Barcelona), Spain}
\footnotetext[23]{Dipartimento di Fisica Sperimentale
dell'Universit\`a di Torino, I-10125 Torino, Italy}
\footnotetext[24]{Institut de Physique Nucleaire de Lyon,
IN2P3-CNRS, Universite Lyon I, F-69622 Villeurbanne, France}
\footnotetext[25]{Funded by the German Federal Minister for Research
and Technology (BMBF) under contract 056SI74}
\footnotetext[26]{University College Dublin School of Physics,
Belfield, Dublin 4, Ireland}
\footnotetext[27]{Centro de Investigaciones Energeticas
Medioambientales y Tecnologicas, E-28040 Madrid, Spain}
\footnotetext[28]{Funded by the Austrian Ministry for Traffic and
Research under the contract GZ 616.360/2-IV GZ 616.363/2-VIII, and
by the Fonds f\"ur Wissenschaft und Forschung FWF Nr.~P08929-PHY}

%%%%%%%%%%%%%%%%%%%%%%%%%%%%%%%%%%%%%%%%%%%%%%%%%%%%%%
\newpage
%\begin{linenumbers}

%%%%%%%%%%%%%%%%%%%%%%%%%%%%%%%%%%%%%%%
\section*{Introduction}

The flavour-changing neutral current decays
$K^\pm\to\pi^\pm\ell^+\ell^-$ (denoted $K_{\pi\ell\ell}$ below,
$\ell=e,\mu$), induced at the one-loop level in the Standard Model
(SM), are well suited to explore its structure and, possibly, its
extensions. The rates of these transitions are dominated by the
long-distance contributions involving one photon exchange. They have
been described in the framework of Chiral Perturbation Theory
(ChPT)~\cite{ek87} in terms of a vector interaction form factor
(which characterizes the dilepton invariant mass spectrum)
determined by experimental measurements. Several models for form
factor have been proposed~\cite{da98,fr04,du06}.

The first experimental observation of the $K_{\pi\mu\mu}^+$ process
was published by the BNL E787 collaboration in 1997~\cite{ad97}. It
was followed by a BNL E865 measurement~\cite{ma00} which established
the vector nature of the decay, and found the form factor and the
decay rate to be in agreement with the expectation based on the
earlier $K_{\pi ee}$ measurements~\cite{al92,ap99}. The most
stringent upper limit on the rate of the lepton number violating
$K^+\to\pi^-\mu^+\mu^+$ decay also comes from the E865
experiment~\cite{ap00}. Later the HyperCP experiment analysed
samples of both $K^+_{\pi\mu\mu}$ and $K^-_{\pi\mu\mu}$
decays~\cite{pa02}, which, in addition to decay rate measurements,
allowed setting a limit on the CP violating rate asymmetry. The
total $K_{\pi\mu\mu}$ sample collected by the three experiments
amounts to $\sim700$ candidates.

A new measurement of the $K_{\pi\mu\mu}^\pm$ decay based on the data
collected by the NA48/2 experiment at the CERN SPS in 2003--2004 is
reported in this letter. The event sample is $\sim 4.5$ times larger
than the total world sample, and has low background contamination,
allowing form factor, rate and asymmetry measurements at an improved
precision.

%%%%%%%%%%%%%%%%%%%%%%%%%%%%%%%%%%%%%%%%%%%%%%%%%
\section{The NA48/2 experiment}

The NA48/2 experiment, specifically designed for charge asymmetry
measurements~\cite{ba07}, uses simultaneous $K^+$ and $K^-$ beams
produced by 400 GeV/$c$ primary SPS protons impinging on a beryllium
target. Beam particles with momentum $(60\pm3)$ GeV/$c$ (r.m.s.) are
selected by an achromatic system of four dipole magnets with zero
total deflection (`achromat'), which splits the two beams in the
vertical plane and then recombines them on a common axis. The beams
pass through momentum defining collimators and a series of four
quadrupoles designed to focus the beams at the detector entrance
plane. Finally the two beams are again split in the vertical plane
and recombined in a second achromat.

The beams enter the fiducial decay volume housed in a 114 m long
cylindrical vacuum tank with a diameter of 1.92 m upstream,
increasing to 2.4 m downstream. Both beams follow the same path in
the decay volume: their axes coincide within 1~mm, while the
transverse size of the beams is about 1~cm. With $7\times 10^{11}$
protons incident on the target per SPS spill of about $4.8$~s
duration, the positive (negative) beam flux at the entrance of the
decay volume is $3.8\times 10^7$ ($2.6\times 10^7$) particles per
pulse, of which $5.7\%$ ($4.9\%$) are $K^+$ ($K^-$). The $K^+/K^-$
flux ratio is 1.79. The fraction of beam kaons decaying in the
vacuum tank at nominal momentum is $22\%$.

A detailed description of the NA48 detector can be found in
Ref.~\cite{fa07}. The decay volume is followed by a magnetic
spectrometer housed in a tank filled with helium at nearly
atmospheric pressure, separated from the vacuum tank by a thin
($\sim 0.4\%X_0$) $\rm{Kevlar}\textsuperscript{\textregistered}$
window. A thin-walled aluminium beam pipe of 16~cm outer diameter
traversing the centre of the spectrometer (and all the following
detector elements) allows the undecayed beam particles and the muon
halo from decays of beam pions to continue their path in vacuum. The
spectrometer consists of four octagonal drift chambers (DCH)
composed of eight planes of sense wires: DCH1, DCH2 located
upstream, and DCH3, DCH4 downstream of a dipole magnet. The magnet
provides a horizontal transverse momentum kick $\Delta p=120~{\rm
MeV}/c$ for charged particles. The spatial resolution of each DCH is
$\sigma_x=\sigma_y=90~\mu$m. The momentum resolution of the
spectrometer is $\sigma_p/p = (1.02 \oplus 0.044\cdot p)\%$ ($p$ in
GeV/$c$).

A plastic scintillator hodoscope (HOD) used to produce fast trigger
signals and to provide precise time measurements of charged
particles is placed after the spectrometer. The HOD has a regular
octagonal shape, and consists of a plane of vertical strip-shaped
counters followed by a plane of horizontal ones (128 counters in
total).

The HOD is followed by a liquid krypton electromagnetic calorimeter
(LKr) used for particle identification in the present analysis. It
is an almost homogeneous ionization chamber with an active volume of
7 m$^3$ of liquid krypton, segmented transversally into 13248
projective cells, approximately 2$\times$2 cm$^2$ each, $27X_0$ deep
and with no longitudinal segmentation. The transverse sizes of the
HOD, DCHs and LKr are about 2.4~m.

A muon detector (MUV) essential for muon identification in the
present analysis is located further downstream. The MUV is composed
of three planes of plastic scintillator strips (aligned horizontally
in the first and last planes, and vertically in the middle plane)
read out by photomultipliers at both ends. Each strip is 2.7~m long
and 1~cm thick. The widths of the strips are 25~cm in the first two
planes, and 45~cm in the third plane. The MUV is also preceded by a
hadronic calorimeter (not used for the present measurement), which
is an iron-scintillator sandwich with a total iron thickness of
1.2~m. Each MUV plane is preceded by an additional 0.8~m thick iron
absorber.

A dedicated two-level trigger has been designed for collection of
three-track decays. A description of the trigger algorithm and the
sources of its inefficiency, which is typically at the $10^{-3}$
level and is therefore negligible for the present analysis, can be
found in Ref.~\cite{ba07}.

A detailed GEANT3-based~\cite{geant} Monte Carlo (MC) simulation
which includes full detector geometry and material description,
stray magnetic fields, DCH local inefficiencies and misalignment,
detailed simulation of the kaon beam line, and time variations of
the above throughout the running period is used to compute the
acceptances for signal, normalisation, and background channels.

%%%%%%%%%%%%%%%%%%%%%%%%%%%%%%%%
\section{Event selection}
\label{sec:sel}

\noindent The $K_{\pi\mu\mu}$ rate is measured relative to the
abundant $K^\pm\to\pi^\pm\pi^+\pi^-$ normalisation channel (denoted
$K_{3\pi}$ below). The $K_{\pi\mu\mu}$ and $K_{3\pi}$ samples are
collected concurrently using the same trigger logic. The fact that
the $\mu^\pm$ and $\pi^\pm$ masses are close ($m_\mu/m_\pi=0.76$)
results in similar topologies of the signal and normalisation final
states. This leads to first order cancellation of the systematic
effects induced by imperfect kaon beam description, local detector
inefficiencies, and trigger inefficiency.

\vspace{2mm}

\noindent {\bf Selection conditions}

\vspace{2mm}

\noindent Three-track vertices (compatible with either
$K_{\pi\mu\mu}$ or $K_{3\pi}$ decay topology) are reconstructed by
extrapolation of track segments from the spectrometer upstream into
the decay volume, taking into account the measured Earth's magnetic
field, stray fields due to magnetization of the vacuum tank, and
multiple scattering.

The selection procedures for the $K_{\pi\mu\mu}$ and $K_{3\pi}$
modes have a large common part: the presence of a vertex satisfying
the following criteria is required.
\begin{itemize}
\item The vertex longitudinal position is within the fiducial decay volume
(i.e. downstream the final collimator).
\item The vertex tracks are required to be consistent in time
(within a 10~ns time window), consistent with the trigger time, and
to be in DCH, HOD, LKr and MUV geometric acceptances. Track momenta
are required to be above 10~GeV/$c$ to ensure high muon
identification efficiency. Track separations are required to exceed
2~cm in the DCH1 plane to suppress photon conversions, and 20~cm in
the LKr front plane to minimize particle misidentification due to
shower overlaps.
\item The total charge of the three tracks is $Q=\pm1$.
\item The total momentum of the three tracks $|\sum\vec p_i|$ is
consistent with the beam nominal range: $(54;66)~{\rm GeV}/c$.
\item The total transverse momentum of the three tracks with respect to
the mean beam direction (which is precisely measured using the
$K_{3\pi}$ sample) is $p_T^2<0.5\times 10^{-3}~({\rm GeV}/c)^2$.
\end{itemize}
If several vertices satisfy the above conditions, the one with the
lowest fit $\chi^2$ is considered. The $K_{\pi\mu\mu}$ candidates
are then selected using the following particle identification and
kinematic criteria.
\begin{itemize}
\item The vertex is required to be composed of one $\pi^\pm$ candidate
(with the ratio of energy deposition in the LKr calorimeter to
momentum measured by the spectrometer $E/p<0.95$, which suppresses
electrons, and no in-time associated hits in the MUV), and a pair of
oppositely charged $\mu^\pm$ candidates (with $E/p<0.2$ and
associated hits in the first two planes of the MUV). The muon
identification efficiency has been measured to be above $98\%$ for
$p>10$~GeV/$c$, and above $99\%$ for $p>15$~GeV/$c$.
\item The invariant mass of the three tracks in the
$\pi^\pm\mu^+\mu^-$ hypothesis lies in the range
$|M_{\pi\mu\mu}-M_K|<8~{\rm MeV}/c^2$, where $M_K$ is the nominal
charged kaon mass.
\end{itemize}
Independently, the following criteria are applied to select the
$K_{3\pi}$ sample.
\begin{itemize}
\item The pion identification criterion described above is applied
to a single pion only, to symmetrize the selection of the signal and
normalisation modes and diminish the corresponding systematic
uncertainties.
\item The invariant mass of the three tracks in the $3\pi^\pm$
hypothesis lies in the range $|M_{3\pi}-M_K|<8~{\rm MeV}/c^2$.
\end{itemize}
No restrictions are applied to the additional energy deposition in
the LKr calorimeter, which decreases sensitivity to accidental
activity.

\vspace{2mm}

\noindent {\bf Signal sample}

\vspace{2mm}

\begin{figure}[tb]
\begin{center}
{\resizebox*{0.5\textwidth}{!}{\includegraphics{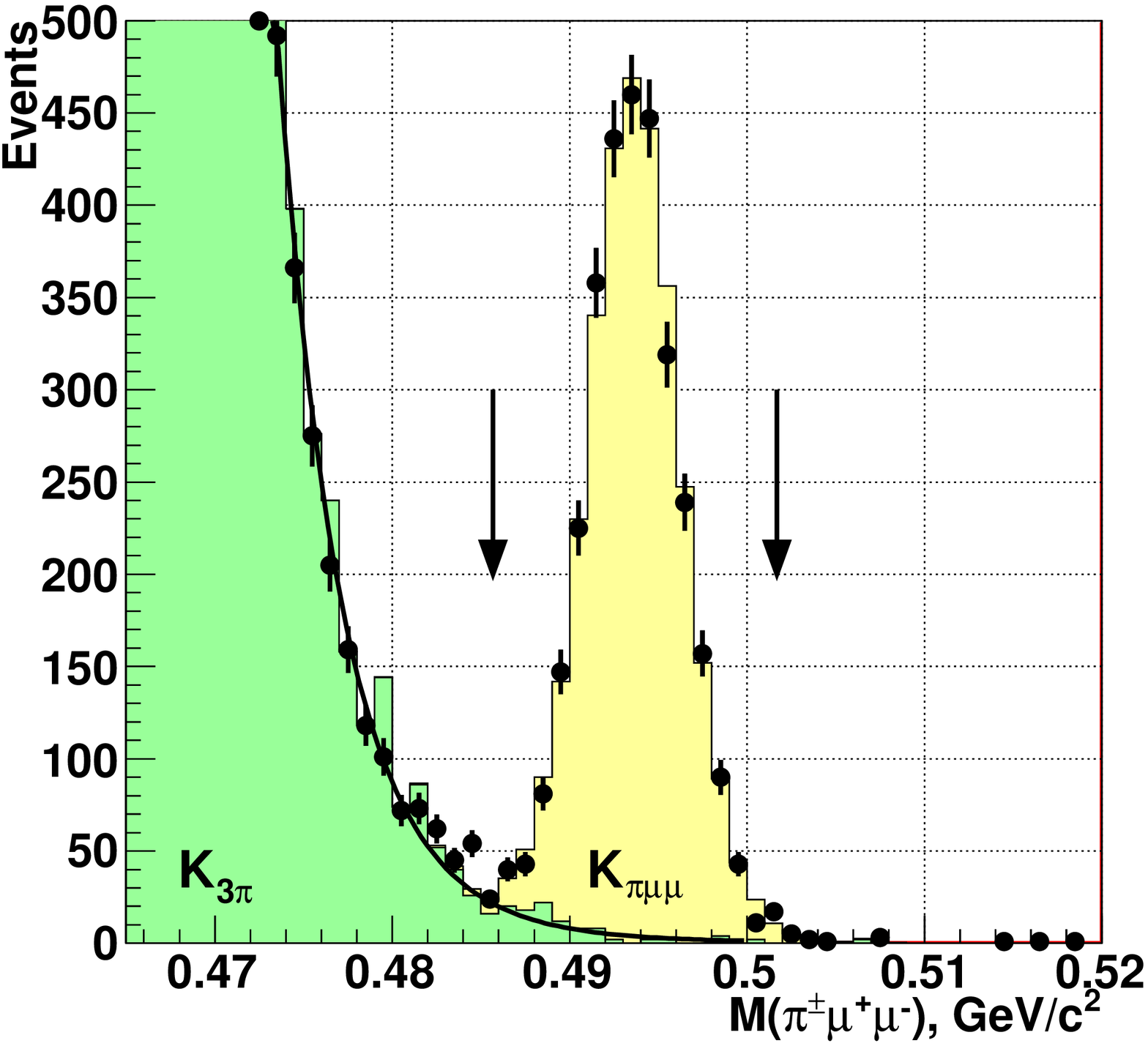}}}%
{\resizebox*{0.5\textwidth}{!}{\includegraphics{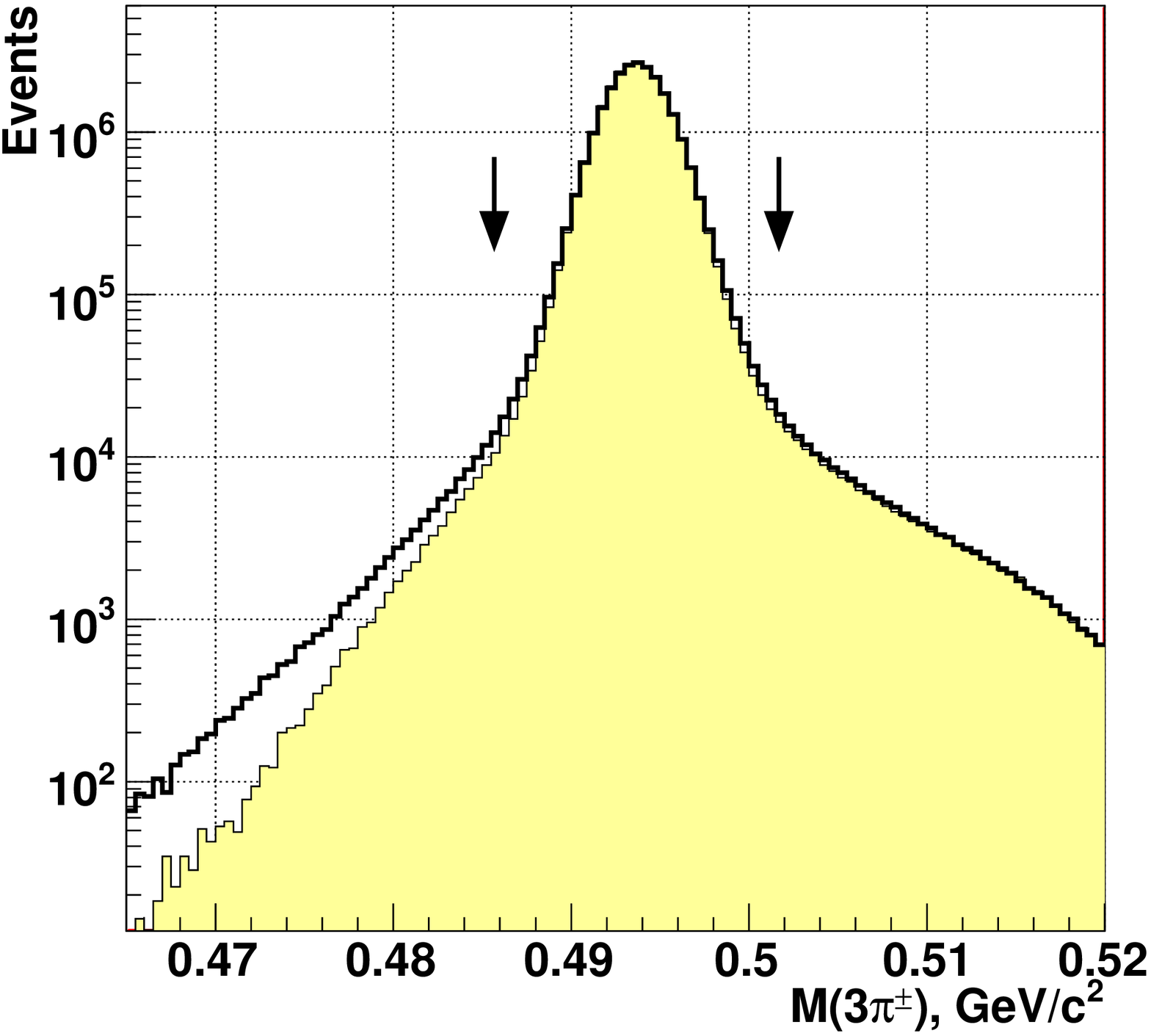}}}
\put(-258,191){\bf\large (a)} \put(-30,191){\bf\large (b)}
\end{center}
\vspace{-14mm} \caption{(a) Reconstructed spectrum of $\pi^\pm
\mu^+\mu^-$ invariant mass: data (dots), $K_{\pi\mu\mu}$ MC
simulation and  $K_{3\pi}$ background estimate with WS events
(filled areas), fit to background with an empirical function (smooth
line). (b) Reconstructed spectrum of $3\pi^\pm$ invariant mass: data
(solid line histogram) and MC simulation (filled area). Signal
regions are indicated with arrows.} \label{fig:mk}
\end{figure}

\noindent The reconstructed $\pi^\pm\mu^+\mu^-$ invariant mass
spectrum is presented in Fig.~\ref{fig:mk}a: a $K_{\pi\mu\mu}$ decay
signal is observed. The number of $K_{\pi\mu\mu}$ candidates in the
signal region is $N_{\pi\mu\mu}=3120$, of which 2003 (1117) are
$K^+$ ($K^-$) candidates. The measured $M_{\pi \mu\mu}$ resolution
is $\sigma_{\pi\mu\mu}=2.5$~MeV/$c^2$, in agreement with MC
simulation.

The $K_{3\pi}$ decay is the only significant background source. It
contributes either via in flight decays of the two pions
($\pi^\pm\to\mu^\pm\nu_\mu$), or via a decay of a single pion and
misidentification of another pion as a muon. Only pion decays
resulting in muons almost collinear to the pion direction, and thus
consistent with a three-track vertex and satisfying the total and
transverse momentum requirements, contribute to background.

\begin{figure}[tb]
\begin{center}
{\resizebox*{0.5\textwidth}{!}{\includegraphics{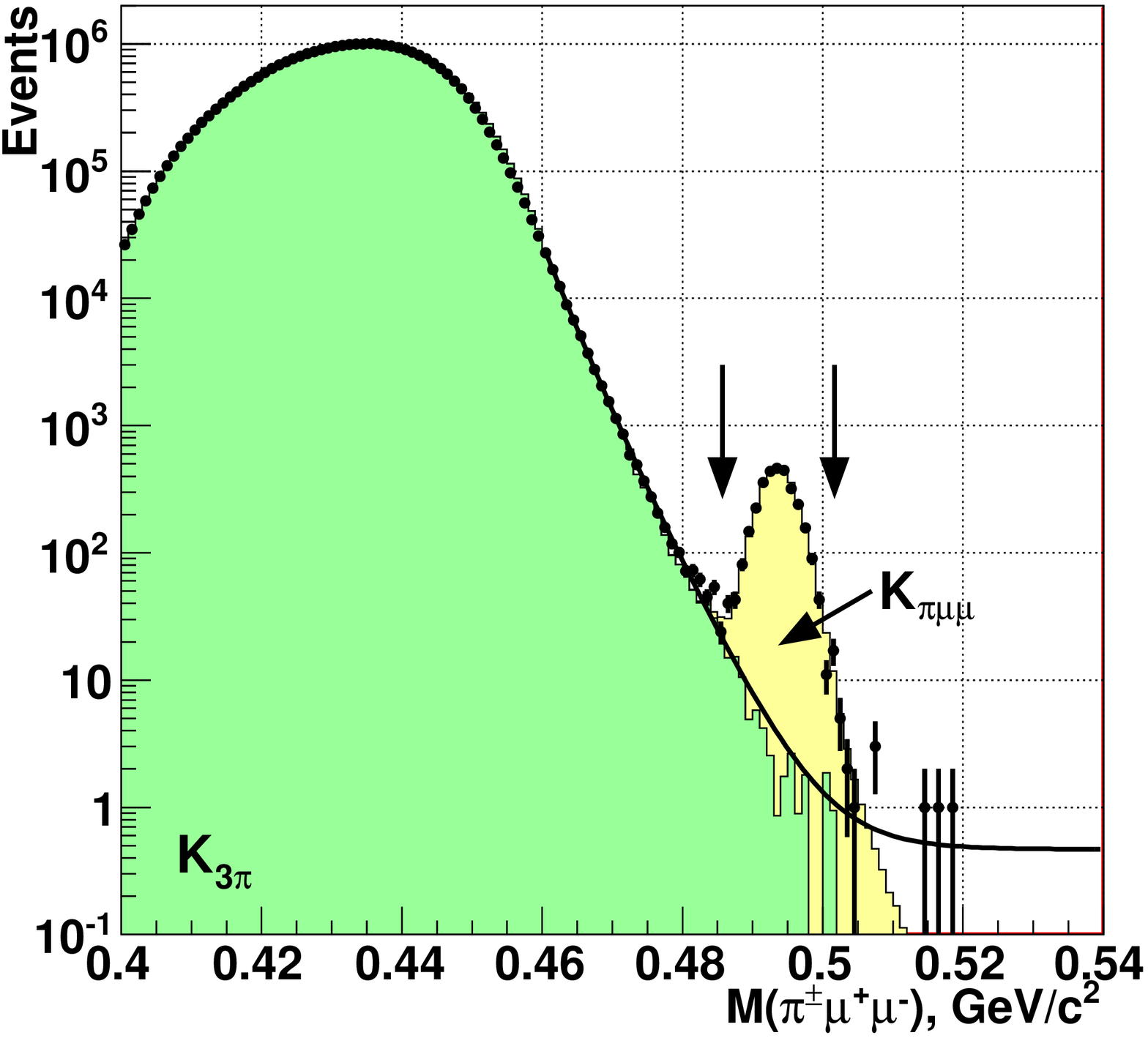}}}%
{\resizebox*{0.5\textwidth}{!}{\includegraphics{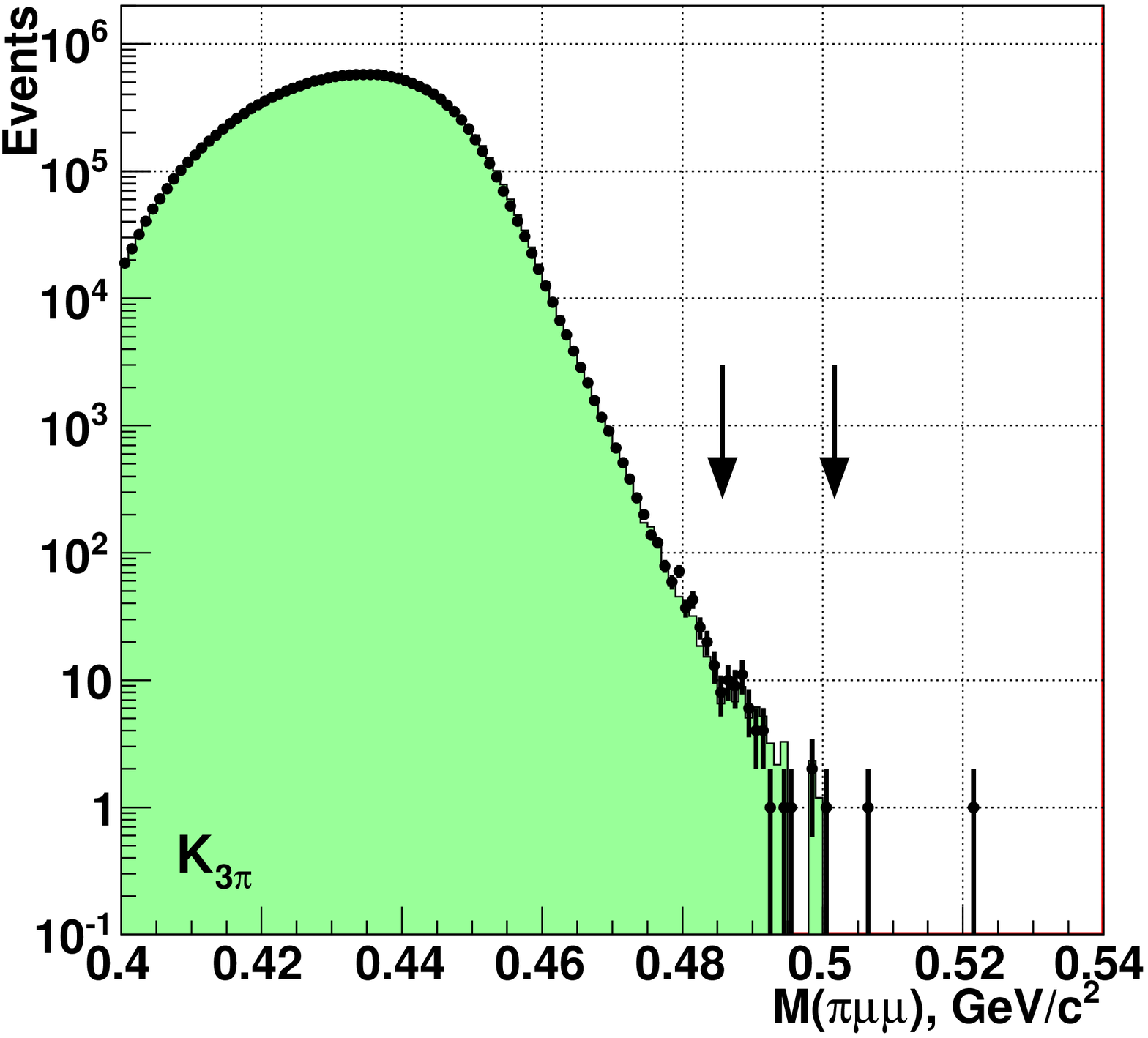}}}
\put(-258,191){\bf\large (a)} \put(-30,191){\bf\large (b)}
\end{center}
\vspace{-14mm} \caption{Reconstructed $M_{\pi\mu\mu}$ spectra of (a)
$K_{\pi\mu\mu}$ and (b) WS candidates: data (dots), $K_{3\pi}$ and
$K_{\pi\mu\mu}$ MC simulations (filled areas); fit to background
using the empirical parameterization as explained in the text (solid
line). The standard signal region is indicated with arrows.}
\label{fig:bkg}
\end{figure}

Three methods of background evaluation are considered.
\begin{itemize}
\item Owing to the symmetry properties of the detector, the
kinematic distribution of the background events is to a good
approximation identical to that of the reconstructed lepton number
violating ``wrong muon sign'' (WS) $\pi^\mp\mu^\pm \mu^\pm$
candidates multiplied by a factor of 2. This observation allows to
estimate the background contamination as $(3.3\pm0.5_{\rm
stat.}\pm0.5_{\rm syst.})\%$, where the quoted statistical
uncertainty is due to the limited number of data WS candidates
($N_{\rm WS}=52$), and the systematic uncertainty is discussed
below.
\item MC simulation of the $K_{3\pi}$ sample leads to a background
estimate of $(2.4\pm0.7)\%$. This method also gives an estimate of
the expected number of WS data events in the signal region: $N_{\rm
WS}^{\rm MC}=52.6\pm19.8$. The quoted uncertainties are systematic
due to the limited precision of MC description of the high-mass
region. They have been estimated from the level of data/MC agreement
in the control reconstructed mass region of (465; 485)~MeV/$c^2$.
\item Fitting the mass spectrum in the region between 460 and
520~${\rm MeV}/c^2$, excluding the signal region between 485 and
502~${\rm MeV}/c^2$, with an empirical function (similar to that
used in the E865 analysis~\cite{ma00,ap00}: a constant plus an
exponentiated cubic polynomial) using the maximum likelihood
estimator and assuming a Poisson probability density in each bin
leads to a background estimate of 3.1\%.
\end{itemize}
The first method is considered the most reliable, and is used in the
subsequent analysis. The $K_{3\pi}$ background estimated with the WS
and extrapolation methods is shown in Fig.~\ref{fig:mk}a. The degree
of agreement of the three background estimation methods is
demonstrated in Fig.~\ref{fig:bkg}.

\begin{figure}[p]
\begin{center}
{\resizebox*{0.5\textwidth}{!}{\includegraphics{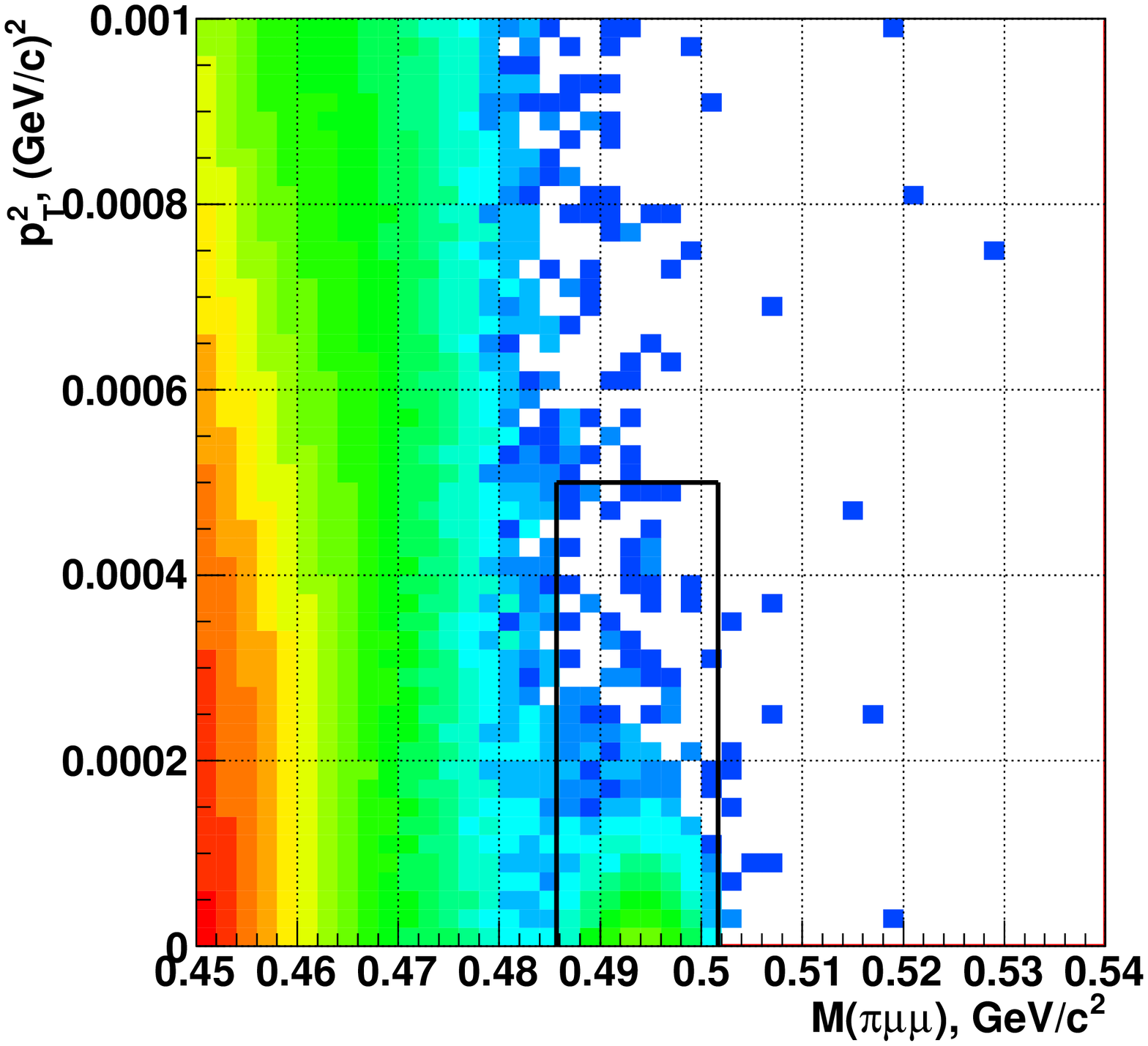}}}%
{\resizebox*{0.5\textwidth}{!}{\includegraphics{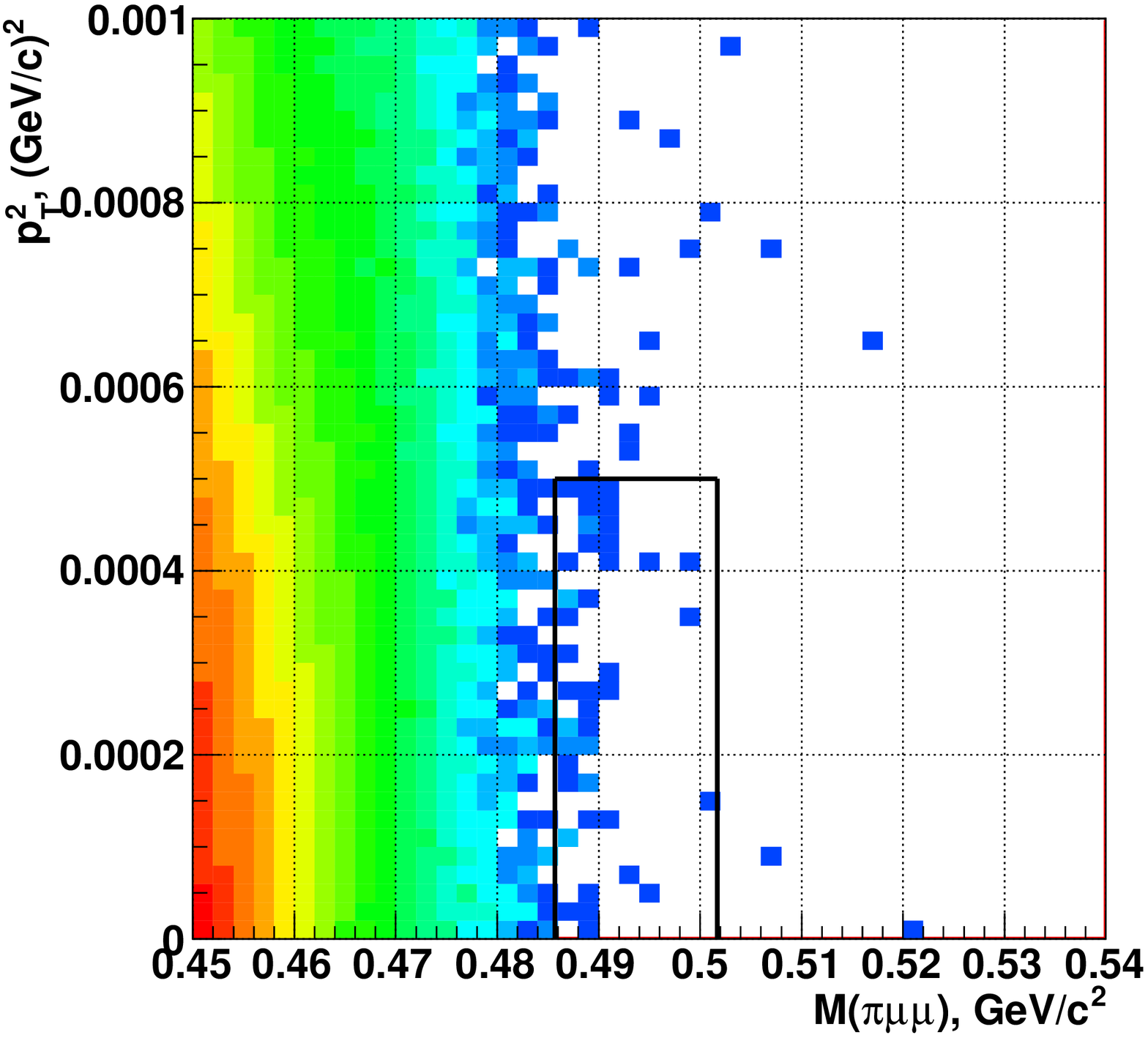}}}
\put(-255,191){\bf\large (a)} \put(-28,191){\bf\large (b)}
\end{center}
\vspace{-14mm} \caption{Distributions of data events in the
$(p_T^2$, $M_{\pi\mu\mu})$ plane for (a) $K_{\pi\mu\mu}$ candidates;
(b) WS events. The signal region is enclosed within rectangles. The
colour coding is in logarithmic scale.} \label{fig:mpt2}
\end{figure}

\begin{figure}[p]
\begin{center}
{\resizebox*{0.5\textwidth}{!}{\includegraphics{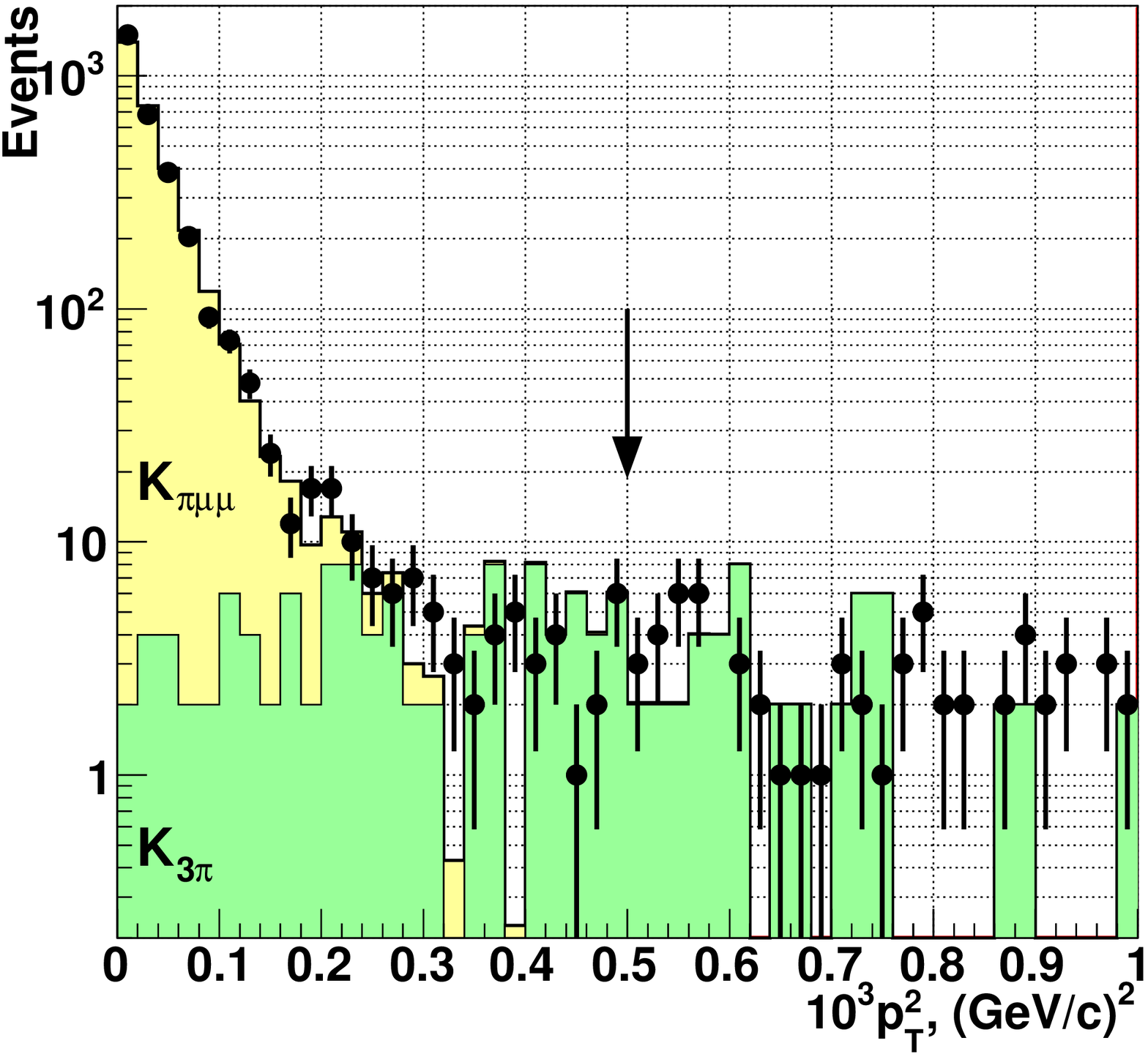}}}
\end{center}
\vspace{-15mm} \caption{Reconstructed $p_T^2$ spectrum of
$K_{\pi\mu\mu}$ candidates: data (dots), $K_{\pi\mu\mu}$ MC
simulation and $K_{3\pi}$ background estimated from WS events
(filled areas). The upper limit of the signal region is indicated
with an arrow.} \label{fig:pt2}
\end{figure}

The distributions of the data $K_{\pi\mu\mu}$ and WS candidates in
the $(p_T^2$; $M_{\pi\mu\mu})$ plane are displayed in
Fig.~\ref{fig:mpt2}. The $p_T^2$ distributions of the data
$K_{\pi\mu\mu}$ candidates, MC $K_{\pi\mu\mu}$ events and $K_{3\pi}$
background evaluated with WS events are shown in Fig.~\ref{fig:pt2}.

\vspace{2mm}

\noindent {\bf Normalisation sample}

\vspace{2mm}

\noindent The normalisation channel $K_{3\pi}$ is well understood in
terms of simulation, being of primary physics interest to
NA48/2~\cite{ba07,slopes}. The reconstructed $3\pi^\pm$ invariant
mass spectrum is presented in Fig.~\ref{fig:mk}b: the non-Gaussian
mass tails are due to $\pi^\pm\to\mu^\pm\nu_\mu$ decays in
flight\footnote{Given that ${\rm BR}(K_{3\pi})/{\rm
BR}(K_{\pi\mu\mu})\sim 10^6$, the filtering of the data stream and
the analysis are performed is such a way that the $K_{3\pi}$
candidates are effectively pre-scaled by a factor of 100, which
strongly reduces the data volume. The shown $K_{3\pi}$ mass plot is
based on the pre-scaled sample.}. The small deficit of MC events at
low $M_{3\pi}$ outside the signal region, explained in part by
radiative $K_{3\pi\gamma}$ decays, is not relevant for the present
analysis. The background contamination is negligible. The number of
$K_{3\pi}$ candidates in the signal region is $2.386\times 10^9$
which, taking into account acceptance, trigger efficiency and
BR~\cite{pdg}, corresponds to a number of kaon decays in the
fiducial volume of $N_K=1.9\times 10^{11}$. The measured $M_{3\pi}$
resolution of $\sigma_{3\pi}=1.7$~MeV/$c^2$ is in agreement with
simulation, and is significantly smaller than $\sigma_{\pi\mu\mu}$
due to the smaller $Q$-value of the $K_{3\pi}$ decay.

%%%%%%%%%%%%%%%%%%%%%%%%%%%%%%%%
\section{Interpretation of the data}

\noindent {\bf Form factor parameterizations} \vspace{2mm}

\noindent The decay is described as proceeding via single virtual
photon exchange, resulting in a spectrum of the
$z=(M_{\mu\mu}/M_K)^2$ kinematic variable sensitive to the form
factor $W(z)$~\cite{ek87}:
\begin{equation}
\frac{d\Gamma}{dz}=\frac{\alpha^2M_K}{12\pi(4\pi)^4}
\lambda^{3/2}(1,z,r_\pi^2)\sqrt{1-4\frac{r_\mu^2}{z}}
\left(1+2\frac{r_\mu^2}{z}\right)|W(z)|^2, \label{theory}
\end{equation}
where $r_\mu=m_\mu/M_K$, $r_\pi=m_\pi/M_K$, and
$\lambda(a,b,c)=a^2+b^2+c^2-2ab-2ac-2bc$. The two-dimensional decay
probability, which is used for evaluation of the geometrical
acceptance with MC simulation, can be found for instance in
Ref.~\cite{du06}. The decay density is corrected by the Coulomb
factor
\begin{equation}
\Omega_C(\beta_{ij})=\prod_{i,j=1,2,3;~i<j} \frac{2\pi\alpha
Q_iQ_j}{\beta_{ij}}\left(e^\frac{2\pi\alpha
Q_iQ_j}{\beta_{ij}}-1\right)^{-1}, \label{coulomb}
\end{equation}
where $Q_i=\pm1$ are the electric charges of the daughter particles,
$0<\beta_{ij}<1$ are their relative velocities, and $\alpha$ is the
fine structure constant. The relative velocities depend on invariant
masses only; in particular, for the muon pair
$\beta_{\mu\mu}^2=1-[2r_\mu^2/(z-2r_\mu^2)]^2$. The following models
of the form factor $W(z)$ are considered.
\begin{enumerate}
\item Linear: $W(z)=G_FM_K^2f_0(1+\delta z)$
with free normalisation and slope $(|f_0|,\delta)$. Decay rate and
spectrum are not sensitive to the choice of the sign of $f_0$.
\item Next-to-leading order ChPT~\cite{da98}:
$W(z)=G_FM_K^2(a_++b_+z)+W^{\pi\pi}(z)$ with free parameters
$(a_+,b_+)$ and an explicitly calculated pion loop term
$W^{\pi\pi}(z)$ given in~\cite{da98}.
\item Combined framework of ChPT and large-$N_c$ QCD
(Ref.~\cite{fr04} and Appendix of Ref.~\cite{piee}): the form factor
is parameterized as $W(z)\equiv W(\tilde{\rm w},\beta,z)$ with free
parameters $(\tilde{\rm w},\beta)$.
\item ChPT parameterization~\cite{du06} involving meson form
factors: $W(z)\equiv W(M_a,M_\rho,z)$. The resonance masses ($M_a$,
$M_\rho$) are treated as free parameters.
\end{enumerate}

\begin{figure}[tb]
\begin{center}
{\resizebox*{0.5\textwidth}{!}{\includegraphics{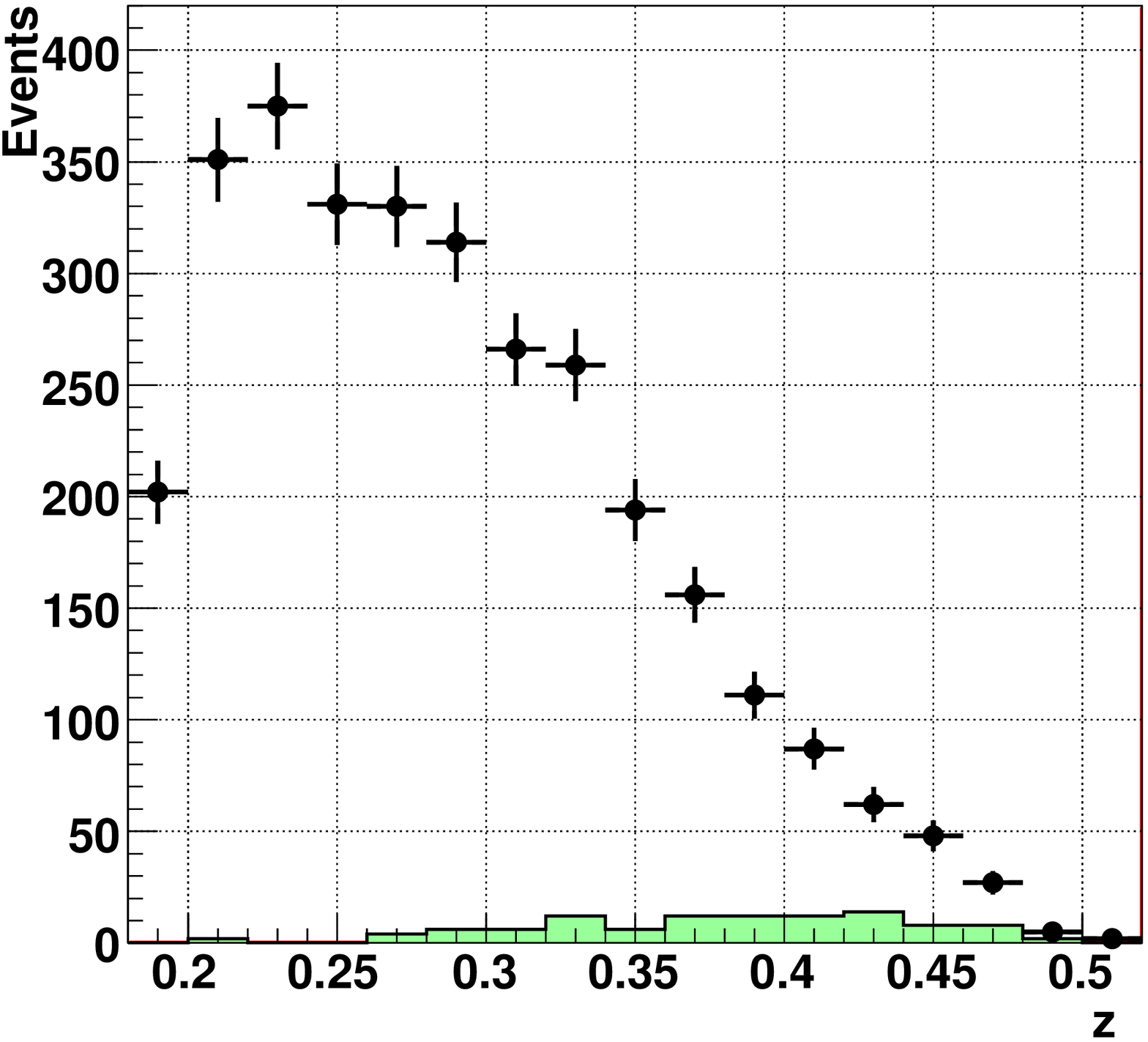}}}%
{\resizebox*{0.5\textwidth}{!}{\includegraphics{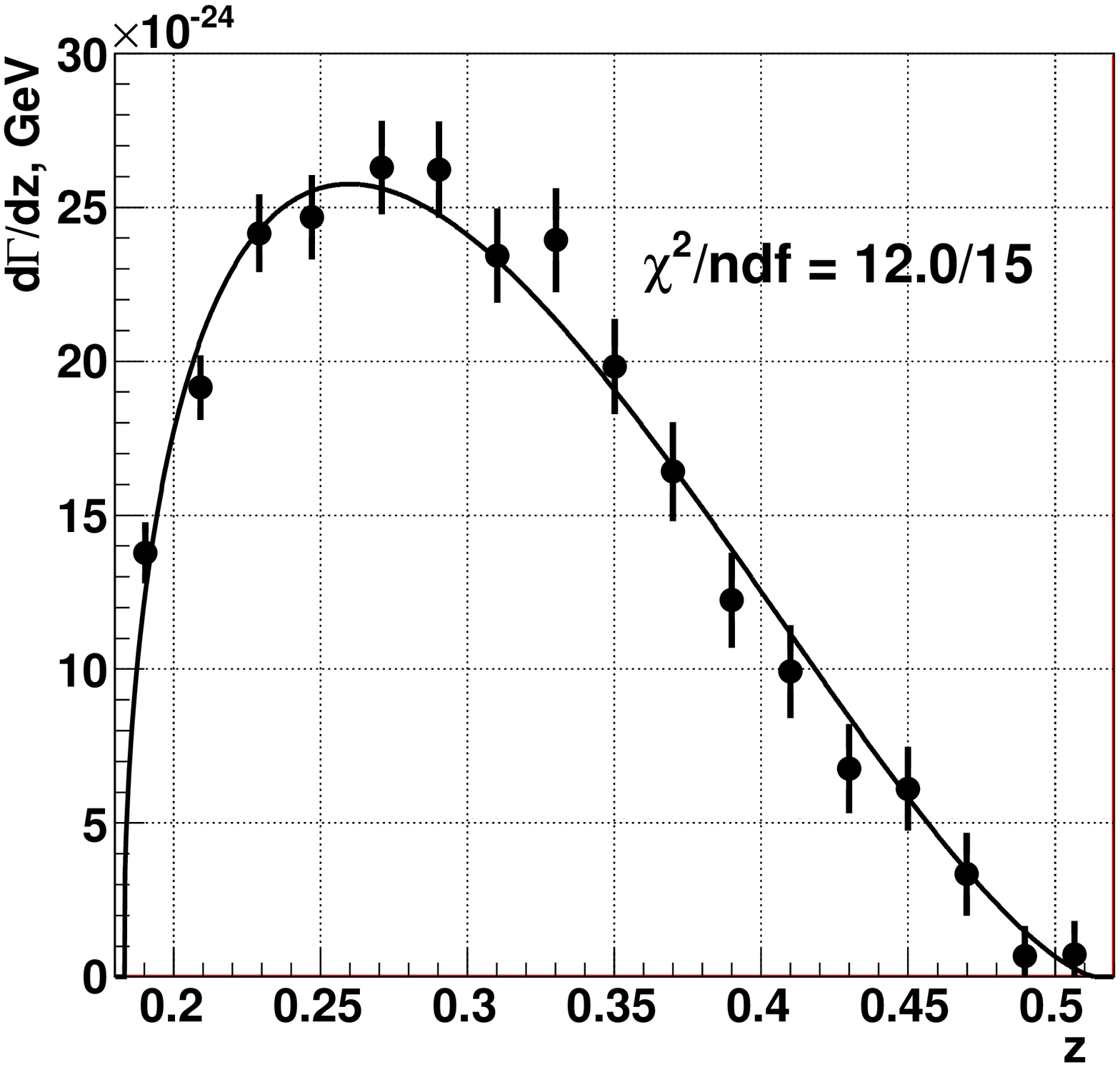}}}
\put(-255,190){\bf\large (a)} \put(-28,190){\bf\large (b)}
\end{center}
\vspace{-14mm} \caption{(a) $z$ spectrum of the selected
$K_{\pi\mu\mu}$ candidates (dots) and $K_{3\pi}$ background
estimated from WS events (filled area). (b) The reconstructed
$d\Gamma_{\pi\mu\mu}/dz$ spectrum fitted to a linear form factor.}
\label{fig:z}
\end{figure}

\vspace{2mm}

\noindent {\bf Form factor measurement} \vspace{2mm}

\noindent The reconstructed $z$ spectrum of the data events is shown
in Fig.~\ref{fig:z}a. The values of $d\Gamma_{\pi\mu\mu}/dz$ in each
$z$ bin, which can be directly compared to the theoretical
expectations (\ref{theory}), are computed as
\begin{equation}
(d\Gamma_{\pi\mu\mu}/dz)_i = \frac{N_i-N^B_i}{N_{3\pi}}\cdot
\frac{A_{3\pi}(1-\varepsilon_{3\pi})}{A_i(1-\varepsilon_i)} \cdot
\frac{1}{\Delta z_i} \cdot \frac{\hbar}{\tau_K} \cdot {\rm
BR}(K_{3\pi}). \label{dgdz}
\end{equation}
Here $N_i$ and $N^B_i$ are the numbers of $K_{\pi\mu\mu}$ candidates
and background events in the $i$-th bin, $N_{3\pi}$ is the number of
selected $K_{3\pi}$ events, $A_i$ (ranging from 5.2\% to 25.2\%) and
$\varepsilon_i$ are the geometrical acceptance and trigger
inefficiency in the $i$-th bin for the signal sample, and
$A_{3\pi}=22.2\%$ and $\varepsilon_{3\pi}$ are those for $K_{3\pi}$
events. Trigger inefficiencies are $\varepsilon\approx0.2\%$,
dominated by timing misalignment and therefore independent of
kinematics and mostly cancelling in~(\ref{dgdz}). The bin widths
$\Delta z_i$ for the $(d\Gamma_{\pi\mu\mu}/dz)_i$ computation are
chosen to be 0.02 for all bins except the first and last ones
limited by the phase space: $4r_\mu^2\le z\le (1-r_\pi)^2$. The
resolution in the $z$ variable is smaller than the bin width,
ranging from 0 at $z=4r_\mu^2$ to 0.007 at $z=(1-r_\pi)^2$. The
external inputs are the kaon lifetime $\tau_K$ and the branching
ratio of the normalisation decay mode ${\rm
BR}(K_{3\pi})$~\cite{pdg}.

The effective $z_i$ values, at which $(d\Gamma_{\pi\mu\mu}/dz)_i$ is
evaluated, are corrected for non-linearity of
$d\Gamma_{\pi\mu\mu}/dz$ following~\cite{la95}. This results in a
significant correction in the first $z$ bin where
$d\Gamma_{\pi\mu\mu}/dz$ has the largest gradient.

The values of $d\Gamma_{\pi\mu\mu}/dz$ vs $z$ and the result of the
fit to the linear form factor are presented in Fig.~\ref{fig:z}b.
The fits to the other models are very similar, and are not shown.

\vspace{2mm}

\noindent {\bf Model-independent branching fraction and asymmetries}
\vspace{2mm}

\noindent The model-independent BR is evaluated by integration of
the spectrum (\ref{dgdz}) normalised to the full $K^\pm$ decay width
$\hbar/\tau_K$. Separate measurements of the BR for $K^+$ and $K^-$
decays allow the evaluation of the CP violating charge asymmetry of
the decay rates: $\Delta(K_{\pi\mu\mu}^\pm)=({\rm BR}^+-{\rm
BR}^-)/({\rm BR}^++{\rm BR}^-)$.

\begin{figure}[tb]
\begin{center}
{\resizebox*{0.5\textwidth}{!}{\includegraphics{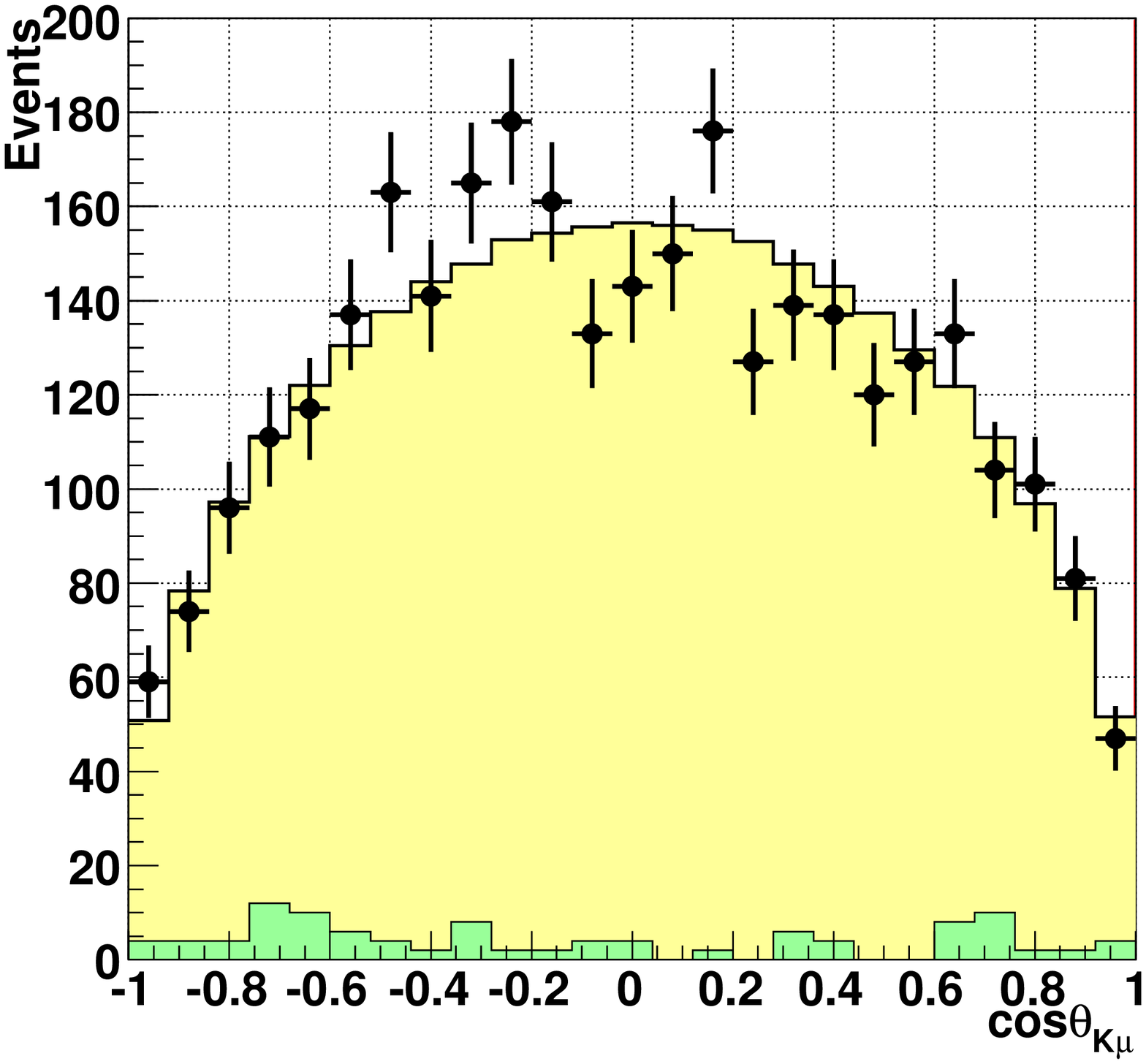}}}%
{\resizebox*{0.5\textwidth}{!}{\includegraphics{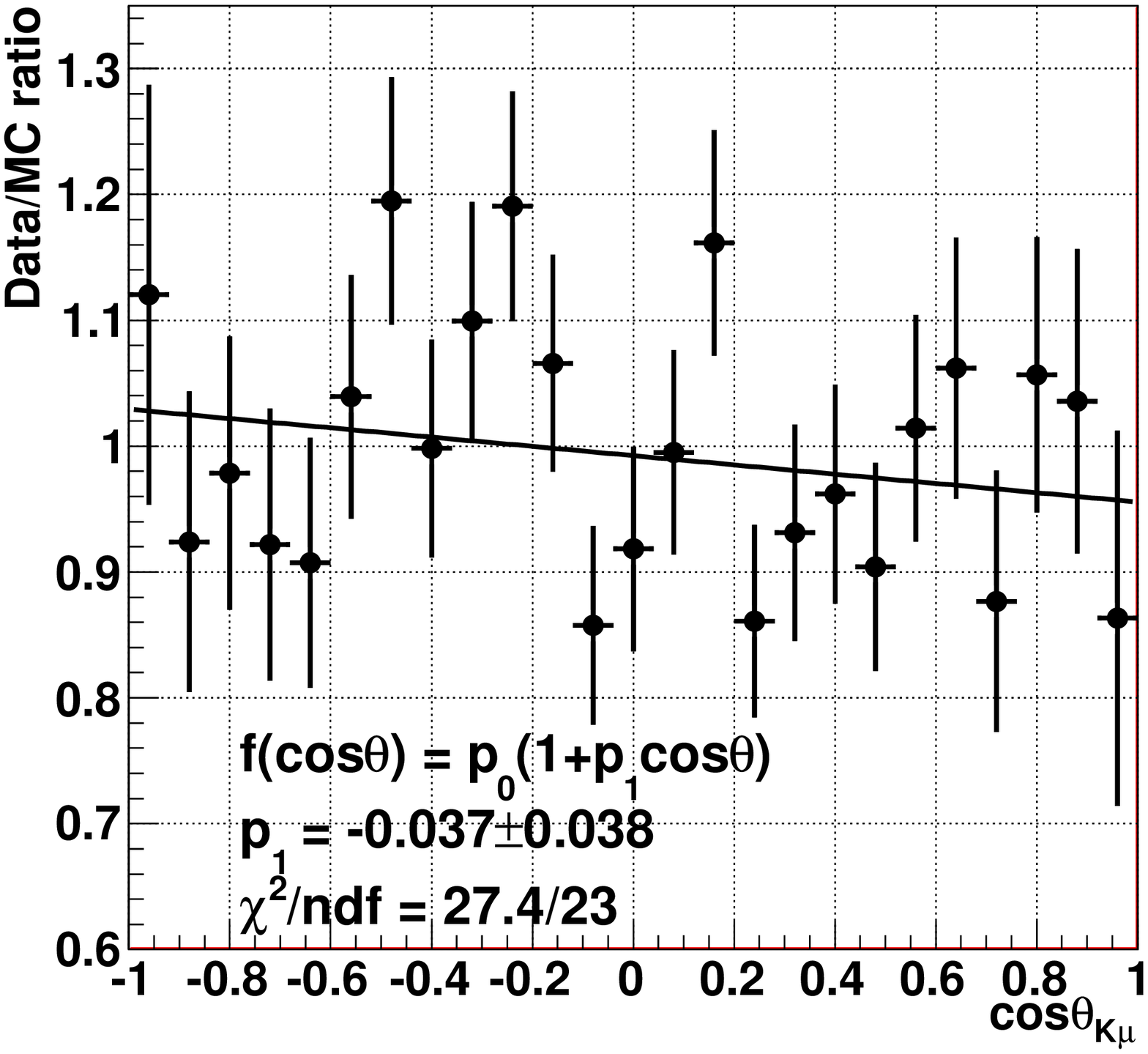}}}
\put(-253,190){\bf\large (a)} \put(-28,190){\bf\large (b)}
\end{center}
\vspace{-14mm} \caption{(a) Reconstructed $\cos\theta_{K\mu}$
spectrum of $K_{\pi\mu\mu}$ candidates: data (dots), $K_{\pi\mu\mu}$
MC simulation and $K_{3\pi}$ background estimated from WS events
(filled areas). (b) Data/MC ratio (background subtracted). The
simulation does not include any forward-backward asymmetry.}
\label{fig:fbasym}
\end{figure}

Another interesting observable is the forward-backward asymmetry in
terms of the angle $\theta_{K\mu}$ between the kaon and opposite
sign muon three-momenta in the dimuon rest frame:
\begin{equation}
A_{FB} = \frac{N(\cos\theta_{K\mu}>0)-N(\cos\theta_{K\mu}<0)}
{N(\cos\theta_{K\mu}>0)+N(\cos\theta_{K\mu}<0)}.
\end{equation}
Non-zero values of the asymmetry can be obtained with the presence
of scalar type interactions. $A_{FB}$ is expected to be sensitive to
non-SM effects, and to be enhanced in the $K_{\pi\mu\mu}$ decay with
respect to the $K_{\pi ee}$ decay~\cite{ch03}.

Due to the limited size of the data sample, only the asymmetry
integrated over the $z$ variable is considered in the present
analysis. The $K^+$ and $K^-$ samples are summed up for this
measurement. The $K_{3\pi}$ background is subtracted using the WS
events. The $\cos\theta_{K\mu}$ spectra (i.e. the sums of
$\cos\theta_{K^+\mu^-}$ and $\cos\theta_{K^-\mu^+}$ distributions)
of the data, simulated $K_{\pi\mu\mu}$ signal and background events,
and the data/MC ratio are displayed in Fig.~\ref{fig:fbasym} (no
forward-backward asymmetry is simulated in MC).

\vspace{2mm}

\noindent {\bf Systematic uncertainties} \vspace{2mm}

\noindent The statistical errors include those due to the numbers of
$K_{\pi\mu\mu}$ candidates and WS data events used to estimate the
background.

The scale factor of 2 applied to WS candidates to model the
background in the $K_{\pi\mu\mu}$ sample has an uncertainty due to
the dependence of $K_{3\pi}$ Dalitz plot density and acceptance on
kinematic variables. MC simulation of the $K_{3\pi}$ decay mode
predicts the scale factor to be $(1.7\pm0.3_{\rm stat.})$. The
stability of the ratio of $M_{\pi\mu\mu}$ spectra of data
$K_{\pi\mu\mu}$ to WS candidates over the control region of $(400;
480)$~MeV/$c^2$ suggests a similar uncertainty of 0.3. A dedicated
study of the $K_{3\pi}$ background to the
$K^\pm\to\pi^+\pi^-e^\pm\nu$ decay leads to a similar
conclusion~\cite{ke4}. The effect of this uncertainty is quoted as
the systematic error due to background subtraction. The background
contamination is thus $(3.3\pm0.5_{\rm stat.}\pm0.5_{\rm
syst.})\%=(3.3\pm0.7)\%$.

Muon identification efficiency measured from a $K^\pm\to\mu^\pm\nu$
sample collected during a special low intensity run with a loose
1-track trigger condition varies from 98.3\% to 99.5\%, depending on
track momentum, in the signal momentum range above 10 GeV/$c$. It is
reasonably well described by MC simulation: the difference between
data and simulated efficiencies is below 0.4\% in the whole analysis
momentum range. This difference is propagated as the corresponding
systematic uncertainty.

Uncertainties due to the pion identification efficiency, which
partially cancel due to symmetrisation of the $K_{3\pi}$ and
$K_{\pi\mu\mu}$ selections, are evaluated by comparing the results
with the pion identification criterion switched on and off for the
$K_{3\pi}$ mode.

The residual effects of the $\sim 0.2\%$ trigger inefficiency are
negligible, as explained above. Uncertainties due to imperfect
simulation of beam momentum and angular distributions are
negligible, as these effects are small and largely cancel due to
similarity of signal and normalisation topologies. Sizable
uncertainties arise from the external input: ${\rm
BR}(K_{3\pi})=(5.59\pm0.04)\%$ is experimentally known with a
limited relative precision of 0.7\%~\cite{pdg}.

Uncertainties on the model parameters and the model-independent BR
are summarised in Table 1. Systematic uncertainties are
significantly smaller than the statistical ones, owing to the
limited size of the data sample. Systematic errors on the
asymmetries $\Delta(K_{\pi\mu\mu}^\pm)$ and $A_{FB}$, which mostly
cancel between the pairs of $K_{\pi\mu\mu}$ samples entering the
asymmetry definitions, are negligible.

\begin{table}[t]
\begin{center}
\caption{Summary of the statistical and systematic uncertainties.}
\vspace{1mm}
\begin{tabular}{rrrrrrr}
\hline Model&Parameter&Statistical&Background&Muon ID&Pion ID&External\\
\hline
(1)&$|f_0|$        &0.039&0.006&    0&    0&0.002\\
   &$\delta$       &0.56 &0.11 & 0.01&    0&    0\\
   \hline
(2)&$a_+$          &0.038&0.006&    0&0.002&0.002\\
   &$b_+$          &0.142&0.027&0.005&0.006&0.005\\
   \hline
(3)&$\tilde{\rm w}$&0.014&0.003&    0&0.001&    0\\
   &$\beta$        &0.61 &0.12 & 0.02& 0.02& 0.02\\
   \hline
(4)&$M_a$/GeV      &0.083&0.016&0.002&0.001&0.001\\
   &$M_b$/GeV      &0.027&0.005&0.001&0.001&0.001\\
\hline
&${\rm BR}\times 10^8$&0.21&0.07&0.04&0.08&0.07\\
\hline
\end{tabular}
\end{center}
\label{tab:syst}
\end{table}

\vspace{2mm} \boldmath \noindent {\bf Search for the lepton number
violating decay $K^\pm\to\pi^\mp\mu^\pm\mu^\pm$} \unboldmath
\vspace{2mm}

\noindent The decay $K^\pm\to\pi^\mp\mu^\pm\mu^\pm$ violating lepton
number by two units can proceed via a neutrino exchange if the
neutrino is a Majorana particle, and the E865 upper limit on its
rate~\cite{ap00} currently provides the strongest constraint on the
effective Majorana neutrino mass $\langle
m_{\mu\mu}\rangle$~\cite{zu00,li00}. This decay has also been
studied in the context of supersymmetric models with $R$-parity
violation~\cite{li00}.

%\begin{figure}[t]
%\vspace{-2mm}
%\begin{center}
%{\resizebox*{0.5\textwidth}{!}{\includegraphics{eps/lfv-fit-bw.eps}}}
%\end{center}
%\vspace{-9mm} \caption{Reconstructed invariant mass distribution of
%the WS candidates, fit to an empirical function (solid line), signal
%corresponding to the 90\% CL upper limit (dashed line). The signal
%region is marked with arrows.} \label{fig:lfv}
%\end{figure}

A new upper limit on $\mathrm{BR}(K^\pm\to\pi^\mp\mu^\pm\mu^\pm)$
can be established by analyzing the WS data mass spectrum shown in
Fig.~\ref{fig:bkg}b. The expected background is estimated by MC
simulation of the $K_{3\pi}$ sample. The Feldman-Cousins
method~\cite{fe98} is employed for confidence interval evaluation;
the systematic uncertainty of the background estimate is taken into
account. The geometrical acceptance is conservatively assumed to be
the smallest of those averaged over the $K_{\pi\mu\mu}$ and
$K_{3\pi}$ samples ($A_{\pi\mu\mu}=15.4\%$ and $A_{3\pi}=22.2\%$).

%%%%%%%%%%%%%%%%%%%%%%%%%%%%%%%%%%%%%%%%%

\section{Results and discussion}

\begin{table}[t]
 \label{tab:results}
\begin{center}
\caption{The measured model parameters, their correlation
coefficients, $\chi^2/\mathrm{ndf}$ of the fits, and the
model-independent BR.}\vspace{1mm}
\begin{tabular}{rrrrrrrrrrrr}
\hline Model (1)$\!\!\!$&&
\multicolumn{4}{l}{$\rho(|f_0|,\delta)=-0.993$}&
\multicolumn{4}{l}{$\chi^2/\mathrm{ndf}=12.0/15$}\\
$|f_0|=\!\!\!$                  &$\!0.470$ &$\!\!\pm\!\!$&$0.039_{\rm stat.}$&$\!\!\pm\!\!$&$0.006_{\rm syst.}$&$\!\!\pm\!\!$&$0.002_{\rm ext.}$&$\!\!=\!\!$&$0.470$&$\!\!\pm\!\!$&0.040\\
$\delta=\!\!\!$                 &$\!3.11$  &$\!\!\pm\!\!$&$0.56_{\rm stat.}$ &$\!\!\pm\!\!$&$0.11_{\rm syst.}$ &             &                  &$\!\!=\!\!$&$3.11$ &$\!\!\pm\!\!$&0.57\\
\hline Model (2)$\!\!\!$&&
\multicolumn{4}{l}{$\rho(a_+,b_+)=-0.976$}&
\multicolumn{4}{l}{$\chi^2/\mathrm{ndf}=14.8/15$}\\
$a_+=\!\!\!$                    &$\!-0.575$&$\!\!\pm\!\!$&$0.038_{\rm stat.}$&$\!\!\pm\!\!$&$0.006_{\rm syst.}$&$\!\!\pm\!\!$&$0.002_{\rm ext.}$&$\!\!=\!\!$&$-0.575$&$\!\!\pm\!\!$&0.039\\
$b_+=\!\!\!$                    &$\!-0.813$&$\!\!\pm\!\!$&$0.142_{\rm stat.}$&$\!\!\pm\!\!$&$0.028_{\rm syst.}$&$\!\!\pm\!\!$&$0.005_{\rm ext.}$&$\!\!=\!\!$&$-0.813$&$\!\!\pm\!\!$&0.145\\
\hline Model (3)$\!\!\!$&& \multicolumn{4}{l}{$\rho(\tilde{\rm
w},\beta)=0.999$}&
\multicolumn{4}{l}{$\chi^2/\mathrm{ndf}=13.7/15$}\\
$\tilde{\rm w}=\!\!\!$          &$\!0.064$ &$\!\!\pm\!\!$&$0.014_{\rm stat.}$&$\!\!\pm\!\!$&$0.003_{\rm syst.}$&             &                  &$\!\!=\!\!$&$0.064$&$\!\!\pm\!\!$&0.014\\
$\beta=\!\!\!$                  &$\!3.77$  &$\!\!\pm\!\!$&$0.61_{\rm stat.}$ &$\!\!\pm\!\!$&$0.12_{\rm syst.}$ &$\!\!\pm\!\!$&$0.02_{\rm ext.}$ &$\!\!=\!\!$&$3.77$ &$\!\!\pm\!\!$&0.62\\
\hline Model (4)$\!\!\!$&&
\multicolumn{4}{l}{$\rho(M_a,M_\rho)=0.999$}&
\multicolumn{4}{l}{$\chi^2/\mathrm{ndf}=15.4/15$}\\
$M_a/({\rm GeV}/c^2)=\!\!\!$    &$\!0.993$ &$\!\!\pm\!\!$&$0.083_{\rm stat.}$&$\!\!\pm\!\!$&$0.016_{\rm syst.}$&$\!\!\pm\!\!$&$0.001_{\rm ext.}$&$\!\!=\!\!$&$0.993$&$\!\!\pm\!\!$&0.085\\
$M_\rho/({\rm GeV}/c^2)=\!\!\!$ &$\!0.721$ &$\!\!\pm\!\!$&$0.027_{\rm stat.}$&$\!\!\pm\!\!$&$0.005_{\rm syst.}$&$\!\!\pm\!\!$&$0.001_{\rm ext.}$&$\!\!=\!\!$&$0.721$&$\!\!\pm\!\!$&0.028\\
\hline
${\rm BR}\times10^8=\!\!\!$     &$\!9.62$  &$\!\!\pm\!\!$&$0.21_{\rm stat.}$ &$\!\!\pm\!\!$&$0.11_{\rm syst.}$ &$\!\!\pm\!\!$&$0.07_{\rm ext.}$ &$\!\!=\!\!$&$9.62$ &$\!\!\pm\!\!$&0.25\\
\hline
\end{tabular}
\end{center}
\end{table}

\begin{figure}[t]
\begin{center}
\vspace{+1mm}
\begin{tabular}{cc}
{\resizebox*{0.4\textwidth}{!}{\includegraphics{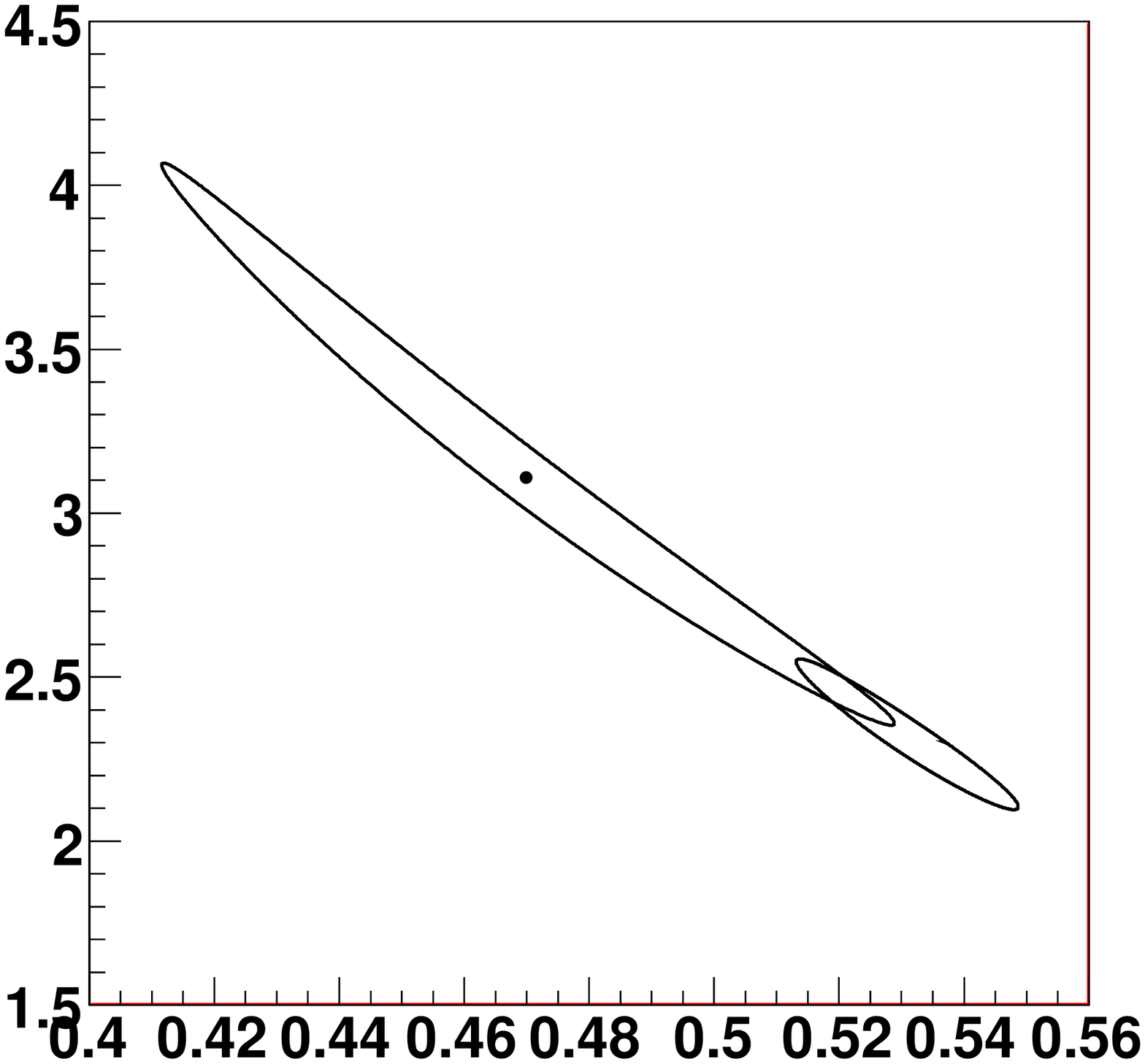}}}
\put(-158,150){\bf\large $\delta$} \put(-33,25){\large $|f_0|$}~&~
{\resizebox*{0.4\textwidth}{!}{\includegraphics{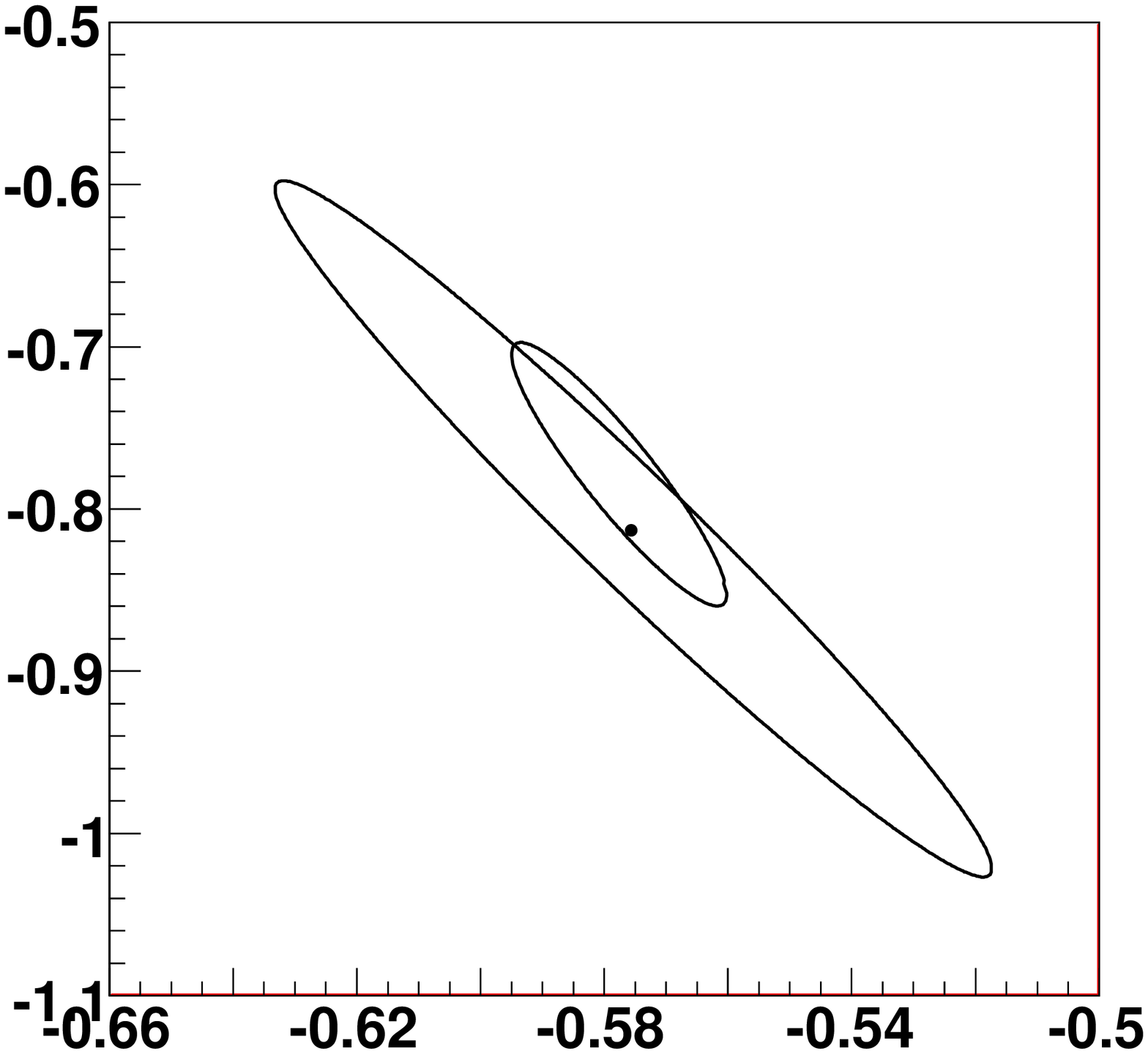}}}
\put(-158,150){\bf\large $b_+$} \put(-29,25){\large $a_+$}\\
{\resizebox*{0.4\textwidth}{!}{\includegraphics{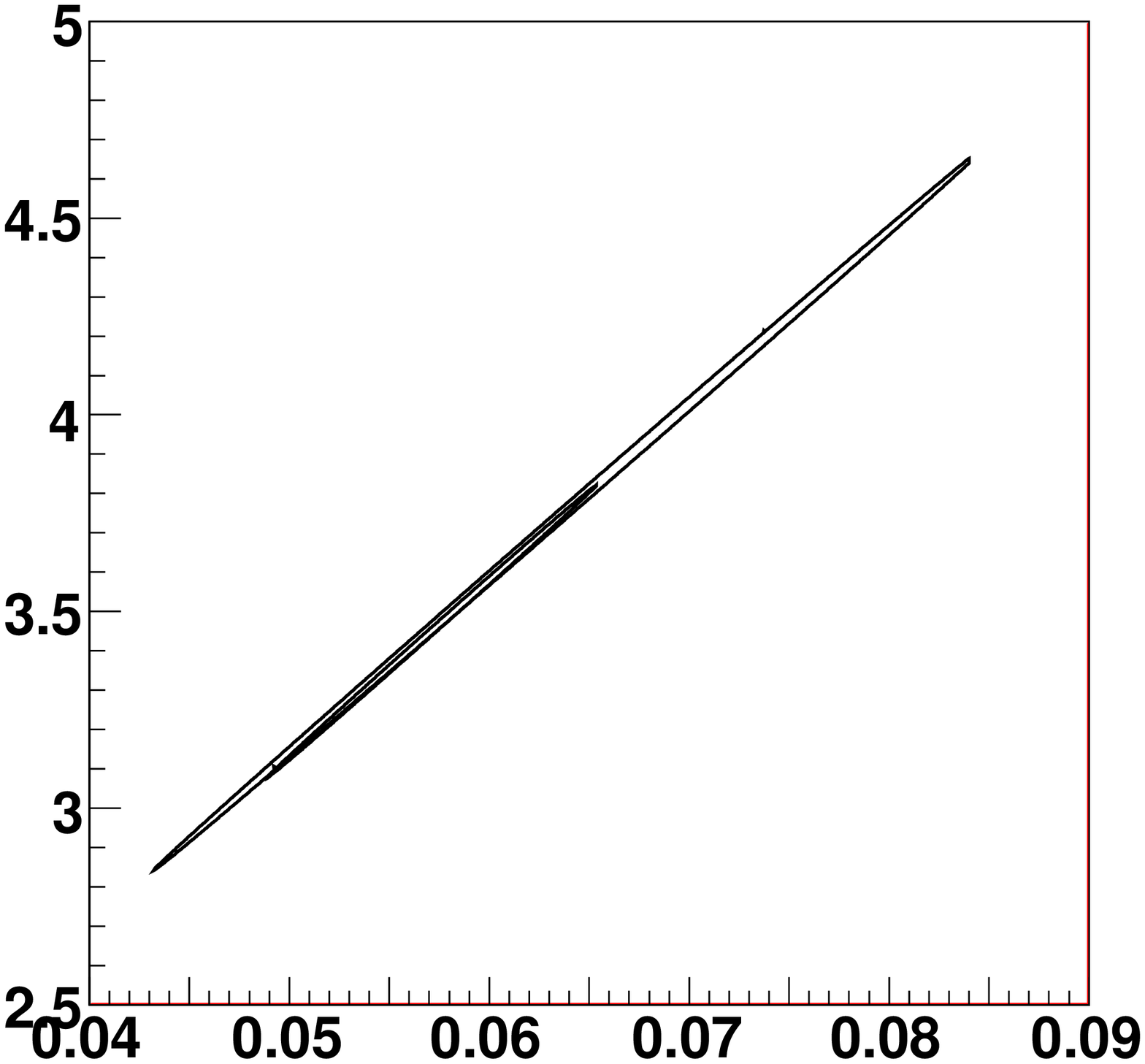}}}
\put(-158,150){\bf\large $\beta$} \put(-25,25){\large $\tilde{\rm
w}$}~&~
{\resizebox*{0.4\textwidth}{!}{\includegraphics{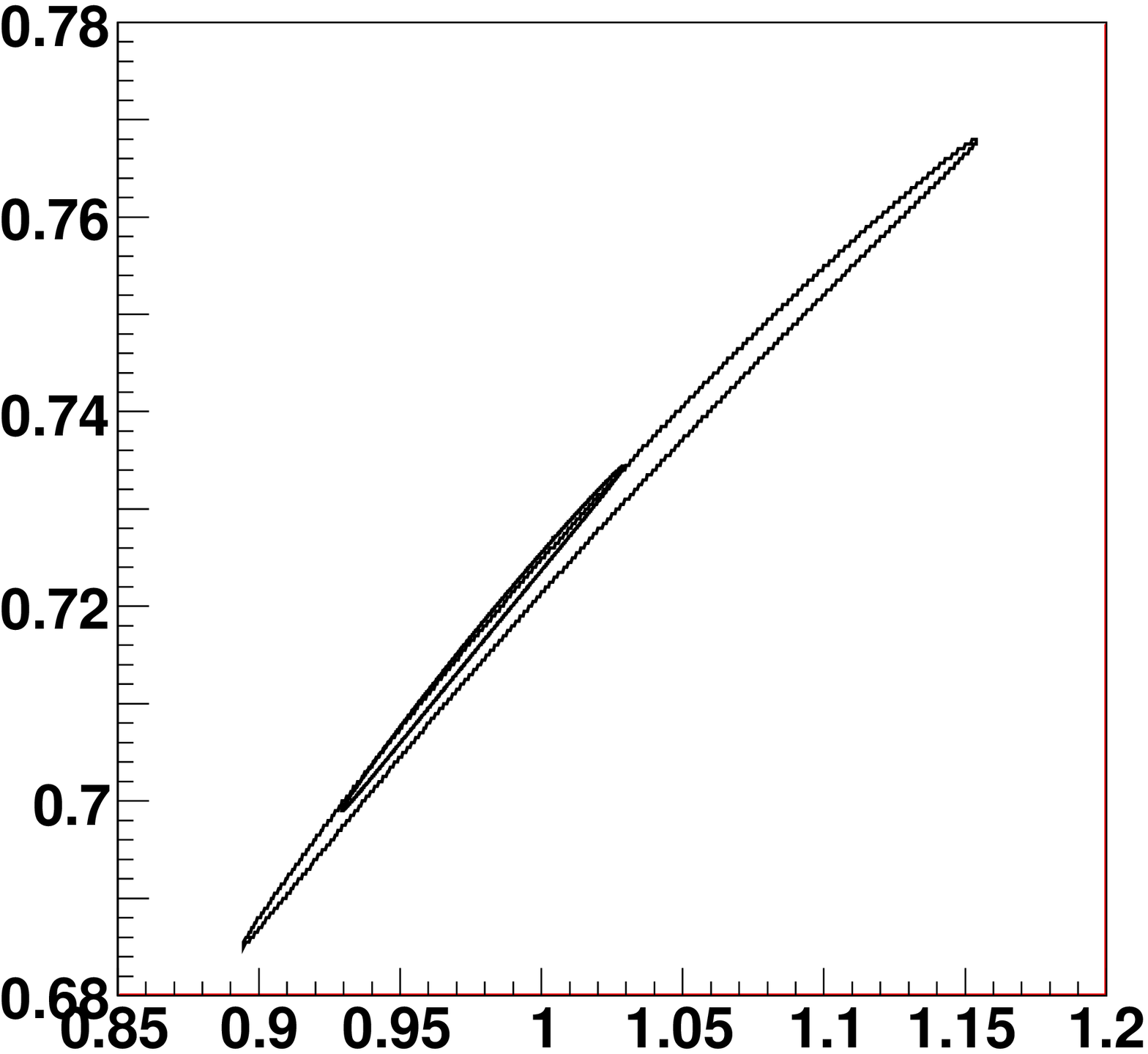}}}
\put(-158,150){\large $M_\rho/({\rm GeV}/c^2)$}
\put(-97,25){\large $M_a/({\rm GeV}/c^2)$}\\
\end{tabular}
\end{center}
\vspace{-9mm} \caption{68\% confidence level contours for the
parameters of the form factor models (larger contours) overlaid with
those from the NA48/2 $K_{\pi ee}$ analysis~\cite{piee} (smaller
contours). The dominant statistical errors only are considered.}
\label{fig:cl}
\end{figure}

\noindent The measured values of the model parameters and the
model-independent BR are presented in Table 2. The overall precision
is limited mainly by the statistical uncertainties. Each of the form
factor models provides a reasonable fit to the data, but the
statistical precision is insufficient to distinguish between the
models. The 68\% confidence level contours for the pairs of
parameters are presented in Fig.~\ref{fig:cl}, overlayed with those
obtained from the analysis of the NA48/2 $K_{\pi ee}$
sample~\cite{piee}.

The measurements of the form factor parameters are summarized in
Table 3. The present measurement is in agreement with the previous
ones based on both $K_{\pi\mu\mu}$~\cite{ma00} and $K_{\pi
ee}$~\cite{al92,ap99,piee} samples. Our measurement of the form
factor linear slope $\delta$ further confirms the contradiction of
the data with the meson dominance models~\cite{li99} which predict
the slope parameter to be in the range from 0.5 to 0.9. The
measurements of the $a_+$ parameter (this result and the $K_{\pi
ee}$ measurements~\cite{ap99,piee}) are in agreement with a
theoretical expectation of
$a_+=-0.6^{+0.3}_{-0.6}$~\cite{br93,pr07}.

\begin{table}[t]
\label{tab:previous}
\begin{center}
\caption{Summary of measurements of the form factor parameters with
their statistical and systematic uncertainties. Some publications
quote statistical errors only, see~\cite{piee} for further
details.}\vspace{1mm}
\begin{tabular}{ccccccc}
\hline
Decay &$K_{\pi ee}^+$ & $K_{\pi ee}^+$ & $K_{\pi ee}^\pm$ & $K_{\pi\mu\mu}^+$ & $K_{\pi\mu\mu}^\pm$\\
Reference &\cite{al92} & \cite{ap99,fr04} & \cite{piee} & \cite{ma00} & this result\\
\hline
$|f_0|$        &             &$\phantom{-}0.533\pm0.012$           &$\phantom{-}0.531\pm0.016$&                                &$\phantom{-}0.470\pm0.040$\\
$\delta$       &$1.31\pm0.48$&$\phantom{-0}2.14\pm0.20\phantom{0}$ &$\phantom{-0}2.32\pm0.18\phantom{0}$&$2.45_{-0.95}^{+1.30}$&$\phantom{-0}3.11\pm0.57\phantom{0}$\\
$a_+$          &             &$-0.587\pm0.010$                     &$-0.578\pm0.016$&                                          &$-0.575\pm0.039$\\
$b_+$          &             &$-0.655\pm0.044$                     &$-0.779\pm0.066$&                                          &$-0.813\pm0.145$\\
$\tilde{\rm w}$&             &$\phantom{-}0.045\pm0.003$           &$\phantom{-}0.057\pm0.007$&                                &$\phantom{-}0.064\pm0.014$\\
$\beta$        &             &$\phantom{-00}2.8\pm0.1\phantom{00}$ &$\phantom{-0}3.45\pm0.30\phantom{0}$&                      &$\phantom{-0}3.77\pm0.62\phantom{0}$\\
$M_a$/GeV     &&&$\phantom{-}0.974\pm0.035$&&$\phantom{-}0.993\pm0.085$\\
$M_{\rho}$/GeV&&&$\phantom{-}0.716\pm0.014$&&$\phantom{-}0.721\pm0.028$\\
 \hline
\end{tabular}
\end{center}
\vspace{-4mm}
\end{table}

The measurements of ${\rm BR}(K_{\pi\mu\mu})$ are summarized in
Table 4. Our measurement is in agreement with those reported
in~\cite{ma00,pa02}, but disagrees with the earliest one~\cite{ad97}
by 4.5 standard deviations. It should be noted that the BR
measurement~\cite{ad97}, obtained under the assumption of the form
factor slope measured with $K_{\pi ee}$~\cite{al92}, is inconsistent
with theoretical expectations based on $K_{\pi ee}$ form factor
measurements assuming that the $K_{\pi\ell\ell}$ decays are
dominated by the $K\to\pi\gamma^*$ form factor~\cite{da98}. The
precision of our BR measurement represents a factor of $\sim3$
improvement with respect to the most precise earlier
measurement~\cite{ma00}.

\begin{table}[t]
\label{tab:previous-br}
\begin{center}
\caption{Summary of ${\rm BR}(K_{\pi\mu\mu})$
measurements.}\vspace{1mm}
\begin{tabular}{cccc}
\hline Reference & Beams & $K_{\pi\mu\mu}$ candidates & ${\rm BR}\times 10^8$\\
\hline
E787~\cite{ad97} & $K^+$   &  207 & $5.0\pm0.4\pm0.6\pm0.7$\\
E865~\cite{ma00} & $K^+$   &  430 & $9.22\pm0.60\pm0.49$\\
HyperCP~\cite{pa02} & $K^\pm$ &  110 & $9.8\pm1.0\pm0.5$\\
NA48/2 & $K^\pm$ & 3120 & $9.62\pm0.21\pm0.11\pm0.07$\\
\hline
\end{tabular}
\end{center}
\vspace{-4mm}
\end{table}

The model-independent branching ratios measured separately for $K^+$
and $K^-$ decays are
\begin{displaymath}
{\rm BR}^+=(9.70\pm0.26)\times10^{-8},~~~{\rm
BR}^-=(9.49\pm0.35)\times10^{-8},
\end{displaymath}
where the quoted uncertainties are statistical only. Neglecting the
systematic uncertainties of ${\rm BR}^+$ and ${\rm BR}^-$ (which are
small compared to the statistical uncertainties, and mostly common
to the $K^+$ and $K^-$ measurements), and the possible charge
asymmetry of $K_{3\pi}$ decay rates which is experimentally
compatible with zero within $2\times10^{-3}$ precision~\cite{fo67},
we measure the charge asymmetry to be
$\Delta(K_{\pi\mu\mu}^\pm)=(1.1\pm2.3)\times 10^{-2}$. This is a
factor of $\sim 5$ improvement in precision with respect to the only
previous measurement~\cite{pa02}, and is compatible with CP
conservation. A limit for the charge asymmetry of
$|\Delta(K_{\pi\mu\mu}^\pm)|<2.9\times10^{-2}$ at 90\% CL can be
deduced from the above value. The experimental precision is far from
the SM expectation $|\Delta(K_{\pi\mu\mu}^\pm)|\sim
10^{-4}$~\cite{da98} and even the SUSY upper limit
$|\Delta(K_{\pi\mu\mu}^\pm)|\sim 10^{-3}$~\cite{me02,da02} for the
CP violating asymmetry.

The forward-backward asymmetry has been measured to be
$A_{FB}=(-2.4\pm1.8)\times 10^{-2}$, where the error is dominated by
the statistical uncertainty. It corresponds to an upper limit of
$|A_{FB}|<2.3\times 10^{-2}$ at 90\% CL. The achieved precision does
not reach the upper limits for the SM contribution via the
two-photon intermediate state
$K^\pm\to\pi^\pm\gamma^*\gamma^*\to\pi^\pm \mu^+\mu^-$~\cite{ga04}
and MSSM contribution~\cite{ch03}, which are both of the order of
$10^{-3}$.

Finally, $N_{\rm WS}=52$ WS data events are observed in the signal
region with the expected background of $N_{\rm WS}^{\rm
MC}=(52.6\pm19.8)$ estimated from MC simulation. Conservatively
assuming the expected background to be $52.6-19.8=32.8$ events to
take into account its uncertainty, this translates into an upper
limit of 32.2 signal events at 90\% CL, leading to an upper limit of
${\rm BR}(K^\pm\to\pi^\mp\mu^\pm\mu^\pm)<1.1\times 10^{-9}$ at 90\%
CL. This is an improvement by almost a factor of 3 with respect to
the best previous limit~\cite{ap00}, allowing a bound on the
effective Majorana neutrino mass of $\langle
m_{\mu\mu}\rangle\lesssim 300~\mathrm{GeV}/c^2$ to be
established~\cite{zu00}.

%%%%%%%%%%%%%%%%%%%%%%%%%%%%%%%%%%%%%%%%%%%%%%
\section*{Conclusions}

From a sample of 3120 $K^\pm\to\pi^\pm\mu^+\mu^-$ decay candidates
with $(3.3\pm0.7)\%$ background contamination, the model-independent
branching fraction has been measured to be ${\rm
BR}=(9.62\pm0.21_{\rm stat.}\pm0.11_{\rm syst.}\pm0.07_{\rm
ext.})\times10^{-8}$, and the form factor which characterizes the
decay has been evaluated in the framework of four models. Upper
limits for the CP violating charge asymmetry and (for the first
time) the forward-backward asymmetry of the decay rate have been
established. An upper limit of $1.1\times 10^{-9}$ for the branching
fraction of the lepton number violating
$K^\pm\to\pi^\mp\mu^\pm\mu^\pm$ decay has been obtained. The
achieved precisions dominate the currently available measurements.

%%%%%%%%%%%%%%%%%%%%%%%%%%%%%%%%%
\section*{Acknowledgements}

It is a pleasure to express our appreciation to the staff of the
CERN laboratory, the technical staff of the participating
laboratories, universities and affiliated computing centres for
their efforts in operation of the experiment and data processing. We
are grateful to Dao-Neng Gao and Gino Isidori for valuable
discussions.

%\end{linenumbers}
%%%%%%%%%%%%%%%%%%%%%%%%%%%%%%%%%%%%%%%%%%%%

\end{document}